\documentstyle[11pt]{article}
\input amssymb.sty
\newcommand{\p}{\partial}

\newcommand{\ti}[1]{\tilde{#1}}

\newcommand{\Om}{\Omega}
\newcommand{\de}{\delta}
\newcommand{\al}{\alpha}
\newcommand{\te}{\theta}

\newcommand{\la}{\lambda}
\newcommand{\si}{\sigma}

\newcommand{\D}{\Delta}
\newcommand{\G}{\Gamma}
\newcommand{\ve}{\varepsilon}
\newcommand{\ep}{\epsilon}
\newcommand{\mat}[4]{\left(\begin{array}{cc}{#1}&{#2}\\{#3}&{#4}
\end{array}\right)}
\newcommand{\oh}{\frac{1}{2}}
\newcommand{\SL}{{\rm SL}(2,{\mathbb C})}
\newcommand{\GL}{{\rm GL}(2,{\mathbb C})}
\newcommand{\SUt}{{\rm SU}(2)}
\newcommand{\SUo}{{\rm SU}(1,1)}

\newcommand{\SLR}{{\rm SL}(2,{\mathbb R})}

\def\f1#1{\frac{1}{#1}}
\def\mC{{\mathbb C}}
\def\mZ{{\mathbb Z}}
\def\mR{{\mathbb R}}
\def\mN{{\mathbb N}}
\def\mM{{\mathbb M}}

\newcommand{\bfe}{{\bf e}}
\newcommand{\clZ}{{\cal Z}}
\newcommand{\clS}{{\cal S}}

\def\bfe{{\bf e}}

\def\bfx{{\bf x}}

\def\bfC{{\bf C}}
\def\bfL{{\bf L}}

\def\bfZ{{\bf Z}}

\def\bfX{{\bf X}}
\def\bfI{{\bf I}}
\def\bfIL{{\bf IL}}

\def\bmu{\bar{\mu}}
\def\bu{\bar{u}}
\def\bz{\bar{z}}
\def\bx{\bar{x}}

\newcommand{\ot}{\otimes}

\newcommand{\ran}{\rangle}
\newcommand{\lan}{\langle}

\newtheorem{predl}{Proposition}[section]
\newtheorem{defi}{Definition}[section]
\newtheorem{rem}{Remark}[section]
\newtheorem{cor}{Corollary}[section]
\newtheorem{lem}{Lemma}[section]

\renewcommand{\theequation}{\thesection.\arabic{equation}}
\newcommand{\beq}[1]{\begin{equation}\label{#1}}
\newcommand{\eq}{\end{equation}}

\begin{document}
\begin{flushright}
ITEP-TH/32-04 \\
\end{flushright}
\vspace{10mm}
\begin{center}
{\Large
dS-AdS structures in the non-commutative Minkowski spaces.}\\
\vspace{5mm}
M.A.Olshanetsky \\
{ ITEP,117259 Moscow,Russia;}\\
{Dept.Math.Sci., Univ.of Aarhus,}\\
{DK 8000 Aarhus C, Denmark}\\
{e-mail olshanet@itep.ru} \\

V.-B.K.Rogov\\
{ MIIT, 103055, Moscow, Russia} \\
{e-mail vrogov@cemi.rssi.ru}\\
\vspace{5mm}
\end{center}
\begin{abstract}
We consider a family of non-commutative 4d Minkowski spaces with the
signature (1,3) and
two types of spaces with the signature (2,2). The Minkowski spaces are
defined by the common
reflection equation and differ in anti-involutions. There exist two
Casimir elements and the fixing of one of them leads to the
non-commutative "homogeneous" spaces
$H_3$, $dS_3$, $AdS_3$ and light-cones. We present the quasi-classical
description
of the Minkowski spaces. There are three compatible Poisson structures -
quadratic,
linear and canonical. The quantization of the former leads to the
considered Minkowski spaces.
We introduce the horospheric generators of the Minkowski spaces. They
lead to the horospheric description
of $H_3$, $dS_3$ and $AdS_3$. The irreducible representations of
Minkowski spaces $H_3$
and $dS_3$ are constructed. We find the eigen-functions of the
Klein-Gordon equation in the terms
of the horospheric generators of the Minkowski spaces. They give rise to
eigen-functions
on the $H_3$, $dS_3$, $AdS_3$ and light-cones.
\end{abstract}

\today
\tableofcontents

\section{Introduction}
\setcounter{equation}{0}

The non-commutative Minkowski
spaces (NCMS) naturally arise in the string theory. The origin of the
non-commutativity is the B-field background of the open string that ends
on a D-brane (for reviews, see \cite{DN,Sz}).
On the other hand a string propagation on curved space-time manifolds such
as AdS$_3$, its Euclidean version H$_3$ and dS$_3$ was investigated in
detail over the last few years. These investigations include, in
particular,
AdS-CFT correspondence and solutions of 2+1 gravity like BTZ black holes..
Non-commutative deformation of AdS(dS)/CFT correspondence was analyzed in
\cite{JR,P,L}.

We consider a particular family of NCMS. Their distinguishing feature is
a natural action of the
quantum Lorentz
group ${\cal U}_q({\rm SL}_2)$. NCMS of the signature $(1,3)$
drew an active attention over the last fifteen years starting with the
first works \cite{LT,W}.
At present their complete classification is known \cite{D,Z,AR,AKR}.
In fact, the spaces we consider here are included in the list of
Ref.\,\cite{AR}. As in Ref.\,\cite{AKR} we fit the commutation relation
on NCMS
in the form of the reflection equation.
We construct irreducible
representations of associative algebra generated by non-commutative
coordinates of NCMS with the signature (1,3) (see also \cite{K}). By
fixing one of Casimir elements we define
the non-commutative Lobachevsky spaces (NCLS) ($H_3$) and the
non-commutative imaginary
Lobachevsky spaces (NCILS) ($dS_3$). These algebras have a natural
description in
terms of the non-commutative analog of the horospheric coordinates.
The Casimir element that determines the homogeneous spaces is one of the
horospheric generators. It allows us to define corresponding homogeneous
spaces in terms of the
rest horospheric generators.

The final goal of this paper is the
solutions of the Klein-Gordon equations on NCMS in terms of the horospheric
coordinates. By analogy with the classical case, the solutions are
products of
$q$-cylindric functions. The reduction of these solutions to NCLS, NCILS
and the
non-commutative cone is straightforward.

To include in the consideration the non-commutative $AdS_3$ we consider
non-commutative deformations of the Minkowski spaces with the signature
$(+,-,+,-)$ and $(+,-,-,+)$. While in the classical case these spaces are
isomorphic since they are governed by the isomorphic groups
$\SLR\oplus\SLR$
and $\SUo\oplus\SUo$ in the
non-commutative case the situation is different. Though
the commutation relations in these algebras are the same,
the algebras are distinguished in different anti-involutions.
Instead of the quantum Lorentz group we have
the action of non-isomorphic Hopf algebras. It is
${\cal U}_q({\rm SL}_2(\mR))\oplus{\cal U}_q({\rm
SL}_2(\mR))$ with $|q|=1$ in the former case and ${\cal
U}_q\SUo\oplus{\cal U}_q\SUo $
and $q\in\mR$ in
the latter. The horospheric description and the solutions of the
Klein-Gordon
equations in the case of the space of type $(+,-,+,-)$ are presented as
inthe
case of the signature (1,3).

The properties of the non-commutative "homogeneous spaces" can
be arranged in the following table.

\bigskip
\begin{center}
\begin{tabular}{|c|c|c|c|}
\hline
Notations & q & Hopf algebra symmetry & NCMS \\
\hline\hline
H$_3$ &$q\in\mC$ & ${\cal U}_q(\SL)$ &$\mM_{\de,q}^{1,3}$\\
\hline
$\bfC_{\de,q}^{1,3}$ &$q\in\mC$ & ${\cal U}_q(\SL)$ &$\mM_{\de,q}^{1,3}$\\
\hline
dS$_3$ & $q\in\mC$ & ${\cal U}_q(\SL)$ &$\mM_{\de,q}^{1,3}$\\
\hline
AdS$_3^\pm$ &$|q|=1$ & ${\cal U}_q(\SLR)\oplus{\cal U}_q(\SLR) $ &
$\mM_q^{2,2}$\\
\hline
$\bfC_q^{2,2}$ & $|q|=1$ & ${\cal U}_q(\SLR)\oplus{\cal U}_q(\SLR) $
&$\mM_q^{2,2}$\\
\hline
${\rm \widetilde{ AdS}}_3^\pm$& $q\in(0,1]$ & ${\cal
U}_q(\SUo)\oplus{\cal U}_q(\SUo)$ & $\widetilde{\mM}_q^{2,2}$\\
\hline
${\rm \widetilde{\bfC}}^{2,2}_q$& $q\in(0,1]$ & ${\cal
U}_q(\SUo)\oplus{\cal U}_q(\SUo)$ & $\widetilde{\mM}_q^{2,2}$\\
\hline
\end{tabular}
\end{center}
\bigskip

We also consider the quasi-classical approximations of NCMS.
In this way we obtain quadratic Poisson algebras with anti-involutions.
They are described by the classical reflection equation. The
Poisson algebras have two classical
Casimir functions. The symplectic leaves of these structures
in the case of the signature (1,3) are the classical
horospheres or spheres of $H_3$ and $dS_3$.
We construct the linear and the canonical Poisson structures that are
compatible with the quadratic one.

\bigskip
\noindent
{\sl Notations.}\\
\noindent
Classical variables are denoted by small letters, while their
non-commutative deformation (quantization) with capital letters.
We do not introduce a special notation for the non-commutative
multiplication.

The coordinates of the Minkowski space $\mM^4=(x_1,x_2,x_3,x_4)$ or
$(y_0,y_1,y_2,y_3)$
we identify with the matrix elements of the matrix $\bfx$
\beq{mx}
\bfx=\mat{x_1}{x_2}{x_3}{x_4}=y_0 Id+\sum_\al \ep_\al y_\al\si_\al\,, ~~
\ep_\al=1\,,~{\rm or}~i\,.
\eq
The choice $1$ or $i$ in front of $y_\al$ defines the signature of the
Minkowski space.
The generators of the non-commutative Minkowski space we also arrange in
the matrix form
\beq{0.1}
\bfX=\mat{X_1}{X_2}{X_3}{X_4}\,.
\eq
The deformation parameter is $q=\exp\te\in (0,1]$,
or $q=\exp i\te\,,$ $(|q|=1)$.

\section{Horospheric coordinates on the classical Minkowski spaces }
\setcounter{equation}{0}

There are two types of Minkowski spaces with the signature $(+,-,-,-)$ and
$(+,+,-,-)$. While the first one allows us to describe the Lobachevsky
space
$H_3$ and the Imaginary Lobachevsky space $dS_3$, the second leads to
$AdS_3$. We will consider them separately.

\subsection{Minkowski space $\mM^{1,3}$ in the horospheric description}

The Minkowski space $\mM^{1,3}$ can be identified with the space of
Hermitian matrices
\beq{m13}
\mM^{1,3}=\{\bfx\in{\rm
Mat}_\mC~|~\bfx^\dag=\bfx\}\,,~~(\bar{x}_1=x_1\,,~\bar{x}_2=x_3\,,~\bar{x}_4=x_4)\,.
\eq
The metric is $ds^2=\det(d\bfx)=dx_1dx_4-dx_2dx_3$.
Another set of coordinates is $y_a\,,$ $(a=0,\ldots 3)$ corresponds to the
choice $\ve_\al=1$ in (\ref{mx})
$$
\bfx=\sum_{a=0}^3y_a\sigma_a\,,~~\sigma_0=Id\,.
$$
It leads to the metric
\beq{met}
ds^2=dy_0^2-\sum_{j=1}^3dy^2_j\,.
\eq

The group $\SL$ is the double covering of the proper Lorentz group
SO$(1,3)$ and
acts on the Minkowski space $\mM^{1,3}=\{y_0,\ldots,y_3\}$ as
\beq{1.0}
\bfx\to g^\dag\bfx g\,,~~~g\in\SL\,,
\eq
where $g^\dag$ is the Hermitian conjugated matrix. The action preserves
\beq{det}
\det\bfx=y_0^2-y_1^2-y_2^2-y_3^2=x_1x_4-x_3x_2
\eq
and thereby the metric (\ref{met}) on $\mM^{1,3}$.

The time-like part $\mM^{1,3+}$ of $\mM^{1,3}$ corresponds to the matrices
with
$\det\bfx>0$, while $\det\bfx<0$ corresponds to the space-like part
$\mM^{1,3-}$. The equation $\det\bfx=0$
selects the light cone
\beq{2b.0}
\bfC^{1,3}=\{\det \bfx=x_1x_4-x_2x_3=0\}\,.
\eq

We introduce {\it the horospheric coordinates}
$\bfx\sim (r,h,z,\bz)$. If $x_1\neq 0$ then
\beq{2.0}
x_1=rh\,,~~x_2=rhz\,,~~x_3=rh\bz\,,
\eq
\beq{2a.0}
x_4=r(h|z|^2+\ve h^{-1})\,.
\eq
Here
$$
z\in\mC\,,~~h\in\mR\setminus 0\,,~
\ve=\pm 1,0\,,~r^2\ve=\det\bfx\,.
$$
and
\beq{hd}
z=x_2x_1^{-1}\,,~\bz=x_3x_1^{-1}\,,~
r=\sqrt{|\det\bfx|}\,,~{\rm for}~ \det\bfx\neq 0\,,
\eq
$$
h=\left\{\begin{array}{lcr}
x_1(|\det\bfx|)^{-\oh} & {\rm for} &\ve=\pm 1\,, \\
x_1 &{\rm for}&\ve=0\,. \\
\end{array}\right.
$$
The case $\ve=1$ corresponds to the time-like part of $\mM^{1,3}$, $\ve=-1$
corresponds to the space-like part and $\ve=0$ to the light-cone
$\bfC^{1,3}$.

The horospheric coordinates on the light-cone $\bfC^{1,3}$ are $(h,z,\bz)$
\beq{2c.0}
x_1=h\,,~x_2=hz\,,~x_3=\bar{x_2}\,,~x_4=h|z|^2\,.
\eq

To describe the case $x_1=0$ we put $\ve=-1$, $h\to 0$, $r<\infty$ and
$z\to\infty$
such that $\lim hz=\exp(it)$, $x_2=r\exp(it)$ and $x_4$ takes an arbitrary
real value.
Thus, the horospheric description has the form
$$
(r,\exp(it),x_4)\,,~~x_2=r\exp(it)\,~x_3=r\exp(-it).
$$

Consider
the commutative algebra $\clS(\mM^{1,3})$ of the Schwartz functions on
$\mM^{1,3}$.
The invariant integral with respect to the $\SL$ action on
$\clS(\mM^{1,3})$
$$
I(f)=\int f(x_1,x_2,x_3,x_4)dx_1dx_2dx_2dx_3dx_4
$$
takes the form in the horospheric coordinates
\beq{int}
I(g)=\int g(z,\bz,h,r)r^3hdrdhdzd\bz \,,~~g\in\clS(\mM^{1,3})\,.
\eq

\subsection{Homogeneous spaces, embedded in $\mM^{1,3}$}

The action of $\SL$ (\ref{1.0}) leads to the foliation of $\mM^{1,3}$.
The orbits are defined by fixing $\det\bfx$.
The quadric
$$
\bfL=\{\det\bfx=r_0^2>0\,,~~x_1>0\}
$$
is the upper sheet of the two-sheeted hyperboloid. It is a model of the
Lobachevsky space, or H$_3$.
The metric on H$_3$ is the restriction of the invariant metric
$dx_1dx_4-dx_2dx_3$ on $r={\rm const}$.
In what follows
we assume $r_0=1$. The horospheric coordinates on $\bfL$ have the
restrictions $h>0$. Since $\SUt$ leaves the point $y_0=1,\,y_\al=0$
the Lobachevsky space is the coset $\bfL\sim\SL/\SUt\sim$SO$(1,3)/$SO$(3)$.

Consider the commutative algebra $\clS(\bfL)$ of
Schwartz functions on $\bfL$. Functions from $\clS(\bfL)$ are infinitely
differentiable with all derivatives tending to zero when
$|z|\to\infty,\,h\to\infty,\,h\to 0$ faster than any power.
Let $I_{r_0^2}$ be the ideal in $\clS(\mM^{1,3})$ generated by
$f(\det\bfx-r^2_0)=0$.
The algebra $\clS(\bfL)$ can be described as the factor-algebra
$\clS(\mM^{1,3})/I_1$
with the additional condition $x_1>0$.

In the similar way we describe the upper sheet of the light-cone
$\bfC^{1,3}$
as $\clS(\mM^{1,3})/I_0$.
The horospheric coordinates(\ref{2.0}) being restricted on $\bfC^{1,3+}$
satisfy the condition $(r=1\,,~h>0\,,~\ve=0)$. $\bfC^{1,3+}$ is the
quotient
$\SL/B_\mC$, where $\ti{B}_\mC$ is the subgroup of the form
$$
\ti{B}_\mC=\left\{\mat{\exp (i\phi)}{w}{0}{\exp
(-i\phi)}\,,~w\in\mC\right\}\,.
$$

The space
$$
\bfIL=\{\det\bfx=-1\}
$$
is called the {\em Imaginary Lobachevsky space}.
The corresponding quadric is $y_0^2-\sum_\al y_\al^2=-1$.
It is the de Sitter space $\bfIL\sim
dS_3=\SL/\SUo\sim$SO$(1,3)/$SO$(1,2)$, since
$$
g^\dag\sigma_3 g=\sigma_3\,,~~~{\rm for}~g\in\SUo\,.
$$

As before,
$\clS(\bfIL)\sim\clS(\mM^{1,3})/I_{-1}$, but in contrast with the $\bfL$ and
$\bfC^+$
the horospheric radius $h$ of $\bfIL$ can take an arbitrary value
$h\in\mR\setminus 0$.

We partially compactify $\bfIL$ with respect to the coordinate
$h$. Two "limiting"
space $ \Xi^\pm =h\to \pm\infty$ are called
{\it absolutes}. It follows from
(\ref{2b.0}) and (\ref{2c.0}) that $\Xi^\pm$ can be
considered as the projectivization of the cone $\bfC^{1,3}$. The both
absolutes
are homeomorphic to $\mC$ and therefore can be compactify to
$\overline{\Xi}^\pm\sim\mC P^1$.
Note, that while $\overline{\Xi}^\pm$ are two components of the boundary of
the $\bfIL$,
$\overline{\Xi}^+$ is the boundary of $\bfC^{1,3+}$ and the $\bfL$.

\subsection{Minkowski space $\mM^{2,2}$ in the horospheric description}

We identify $\mM^{2,2}$ with the space Mat$_\mR(2)$.
It is obtained from $\mM^{1,3}$ by the Wick rotation $y_2\to -iy_2$
\beq{ms22}
\mM^{2,2}=\{\bfx\in{\rm Mat}_\mR(2)\}\,,~~\bfx=y_0
Id+y_1\si_1+iy_2\si_2+y_3\si_3\,.
\eq
The metric on $\mM^{2,2}$ assumes the form
\beq{met1}
ds^2=dx_1dx_4-dx_2dx_3=dy_0^2-dy_1^2+dy_2^2-dy_3^2\,.
\eq
Thus, $\mM^{2,2}$ has the signature $(+-+-)$.
The group
$G=\SLR\oplus\SLR$ acts on $\mM^{2,2}$ as
\beq{ac1}
\bfx\to g_2^{-1}\bfx g_1\,,~~g_k\in\SLR\,.
\eq
The transformed matrix belongs to Mat$_\mR(2)$, since $g_1,g_2\in \SLR$.
Moreover, (\ref{ac1}) preserves $\det\bfx$ and therefore the metric.

The horospheric coordinates take the similar form as for $\mM^{1,3}$
after replacing in (\ref{2.0}), (\ref{2a.0}) $z\to z_1$, $\bz\to z_2$,
$(z_1,\,z_2\in\mR)$
\beq{2.0c}
x_1=rh\,,~~x_2=rhz_1\,,~~x_3=rhz_2\,,
\eq
\beq{2a.0d}
x_4=r(hz_1z_2+\ve h^{-1})\,.
\eq

The invariant integral on $\mM^{2,2}$ is the same as on $\mM^{1,3}$
(\ref{int}).

\subsection{Homogeneous subspaces embedded in $\mM^{2,2}$}

The action of $G=\SLR\oplus\SLR$ on $\mM^{2,2}$ preserves the determinant
$\det\bfx$. We have the two types of nontrivial $G$-orbits.

The first type is AdS$_3$ spaces
\beq{ads}
{\rm AdS}^\pm_3=\{\bfx~|~ \det\bfx=\ve r_0^2\,,~\ve=\pm 1\}\,.
\eq
In the commutative case the spaces AdS$^+_3$ and AdS$^-_3$
are isomorphic.
The spaces AdS$_3^\pm$ is identified with the quotient
$(\SLR\oplus\SLR)/\SLR$..
Then AdS$_3\sim\SLR$.

The second type is cone
$$
\bfC^{2,2}=\{\bfx~|~ \det\bfx=0\}.
$$
It is the quotient $\bfC^{2,2}\sim(\SLR\oplus\SLR)/(\ti{B}$,
where $\ti{B}$ is the subgroup of $\SLR)\oplus\SLR$ preserving a point of
$\bfC^{2,2}$. For example, it can be defined as follows
$$
\ti{B}=\left\{(g_2,g_1)\,,~|~\mat{1}{1}{0}{0}g_1=g_2\mat{1}{1}{0}{0}\right\}\,.
$$

\subsection{Minkowski space $\widetilde{\mM}^{2,2}$ in the horospheric
description}

Consider the Minkowski space $\widetilde{\mM}^{2,2}$ with the signature
$(+--+)$.
Though $\widetilde{\mM}^{2,2}$ is isomorphic to $\mM^{2,2}$, we consider it
separately, since they become non-isomorphic in the non-commutative case.

We identify $\widetilde{\mM}^{2,2}$ with the space
of matrix
\beq{tim}
\widetilde{\mM}^{2,2}=\{\bfx\in{\rm
Mat}_\mC(2)~|~\bx_1=x_4\,,~\bx_2=x_3\}\,.
\eq
It is obtained from $\mM^{1,3}$ by the Wick rotations $y_3\to iy_3$.
$$
\bfx=y_0
Id+y_1\si_1+y_2\si_2+iy_3\si_3\,.
$$
The definition (\ref{tim}) is equivalent to
\beq{tim1}
\widetilde{\mM}^{2,2}=\{\bfx\in{\rm
Mat}_\mC~|~\bfx\si_3\bfx^\dag=\si_3\}\,.
\eq
The isomorphism between $\mM^{2,2}$ (\ref{ms22}) and
$\widetilde{\mM}^{2,2}$ is defined
by the conjugation
$$
\ti{x}=gxg^{-1}\,,~~\ti{x}\in\widetilde{\mM}^{2,2}\,~x\in\mM^{2,2}\,,~~
g=\mat{1}{-i}{1}{i}\,.
$$

It follows from (\ref{tim1}) that the group
$G=\SUo\oplus\SUo$ acts on $\widetilde{\mM}^{2,2}$
\beq{ac}
\bfx\to g_2^{-1}\bfx g_1\,,~~g_k\in\SUo\,.
\eq
It means that $\widetilde{\mM}^{2,2}\sim\SUo\oplus\SUo/\SUo$.
The action (\ref{ac}) preserves $\det\bfx$ and therefore the metric.

The horospheric coordinates on $\widetilde{\mM}^{2,2}$ take the form
\beq{2A.0c}
x_1=\frac{r}{2}(h(1+z_1z_2)+\ve h^{-1}+ih(z_1-z_2))\,,~~x_4=\bx_1\,,
\eq
\beq{2A.0d}
x_2=\frac{r}{2}(h(1-z_1z_2)-\ve h^{-1}-ih(z_1+z_2))\,,~~x_3=\bx_2\,,
\eq
where $z_1,z_2\in\mR$.

We again can determine the homogeneous spaces embedded in
$\widetilde{\mM}^{2,2}$ by fixing $\det\bfx=\ve r^2$.
If $\ve=\pm 1$ and $r=r_0$ we come to $\widetilde{\rm AdS}_3^\pm$
isomorphic to the
defined above. The spaces AdS$_3^\pm$ can be identified with $\SUo$.

\subsection{Short summary}

For completeness we consider
the Euclidean space $\mR^4$ and the embedded sphere $S_3$
$$
\bfx=y_0 Id+i\sum_\al y_\al\si_\al\,, ~~
\det\bfx=y_0^2+\sum_\al y_\al^2\,,
$$
$$
S_3=\{\bfx~|~\det\bfx=r_0\}\sim \SUt\oplus\SUt/\SUt\,.
$$
The horospheric description of $S_3$ does not exists.
We will not generalize $S_3$ to the non-commutative case.
We summarize the structure of the homogeneous spaces in the following
table, that will be generalized to the non-commutative situation.
\bigskip
\begin{center}
\begin{tabular}{|c|c|c|c|}
\hline
Notations & Notations & Coset consruction & Ambient space \\
\hline\hline
\bfL & H$_3$ & $\SL/\SUt$ &$\mM^{1,3}$\\
\hline
$\bfC^{1,3}$ & & $\SL/\ti{B}_\mC$ &$\mM^{1,3}$\\
\hline
\bfIL & dS$_3$ & $\SL/\SUo$ &$\mM^{1,3}$\\
\hline
& AdS$_3$ & $\SLR\oplus\SLR/\SLR$ & $\mM^{2,2}$\\
\hline
$\bfC^{2,2}$ & & $\SLR\oplus\SLR/\ti{B}$ &$\mM^{2,2}$\\
\hline
& AdS$_3$ & $\SUo\oplus\SUo/\SUo$ & $\widetilde{\mM}^{2,2}$\\
\hline
&$S_3$& $\SUt\oplus\SUt/\SUt$ & $\mR^4$\\
\hline
\end{tabular}
\end{center}

\section{Laplace operator and its eigen-functions}
\setcounter{equation}{0}

In this paper we generalize to the noncommutative case the following facts
concerning the eigen-functions of the Laplace operator

The solutions of
the Klein-Gordon equation on $\mM^{4}$
\beq{2.5}
\Delta f_\nu(x_1,x_2,x_3,x_4)=\nu^2 f_\nu (x_1,x_2,x_3,x_4)\,,~~
\Delta=\frac{\p^2}{\p x_1\p x_4}-\frac{\p^2}{\p x_3\p x_2}\,.
\eq
are the exponents
\beq{ex}
f_\nu(x_1,x_2,x_3,x_4)=\exp(\xi x)\,,~(\xi x)=\sum\xi_i x_i\,~~
\nu^2=\xi_1\xi_4-\xi_2\xi_3\,.
\eq
We will consider $\Delta$ and its eigen-functions in the horospheric
coordinates.

\subsection{Scalar fields on $\mM^{1,3}$ in the horospheric coordinates}

The metric on $\mM^{1,3}$ in the
horospheric coordinates takes the form
$$
ds^2=g_{jk}dx_jdx_k=\ve dr^2-\ve r^2h^{-2}dh^2-r^2h^2dzd\bz\,.
$$
Then one can rewrite $\Delta=\f1{(\det g)^\oh}\p_jg^{jk}(\det g)^\oh\p_k$
as
\beq{2.7}
\Delta=r^{-2}\left[h^2\frac{\p^2}{\p h^2}+3\frac\p{\p h}+
4\ve h^{-2}\frac{\p^2}{\p\bz\p z}-r^2\frac{\p^2}{\p r^2}-
3r\frac\p{\p r}\right]
\eq
and we come the eigenvalue problem
\beq{2.10}
r^{-2}\left[h^2\frac{\p^2}{\p h^2}+3\frac\p{\p h}+
4\ve h^{-2}\frac{\p^2}{\p\bz\p z}-r^2\frac{\p^2}{\p r^2}-
3r\frac\p{\p r}\right] f_\nu(\bz,h,z;r)=\nu^2f_\nu(\bz,h,z;r)\,.
\eq
Let $\clZ_\nu(x)$ be a cylindric function. It means that $\clZ_\nu(x)$
is a solution of the equation
$$
\frac{\p^2 \clZ_\nu}{\p x^2}+\f1{x}\frac{\p \clZ_\nu}{\p x }+\left(1-
\frac{\nu^2}{x^2}\right)\clZ_\nu=0\,.
$$
We will prove the non-commutative analog of the following statement
\begin{predl}
The basic harmonics of the eigen-value problem (\ref{2.10}) are
\beq{elhar}
f_\nu(\bz,h,z;r)=r^{-1}h^{-1}\exp(i\mu z+i\bmu\bz)
\clZ_{\al}(r\nu)\clZ_{\al}(2i\ve^\oh|\mu|h^{-1})\,,\ve=\pm 1\,,
\eq
and
\beq{elhar1}
f_\nu(\bz,h,z;r)=h^{\al-1}\exp(i\mu z+i\bmu\bz)\,,~~\ve=0\,, ~~
\al^2=\nu^2+1\,,
\eq
where $\mu,\al\in\mC$.
\end{predl}
{\sl Proof}\\
The variables of the equation (\ref{2.10}) can be separated
\beq{bh}
f_\nu(\bz,h,z;r)= \exp(i\mu z+i\bmu\bz)v_\al(h)\chi_{\nu,\al}(r)\,,
\eq
where $\al^2-1$ is the separation constant
\beq{2.11}
\left[h^2\frac{\p^2}{\p h^2}+3h\frac\p{\p h}+
(-4\ve h^{-2}|\mu|^2-\al^2+1)\right]v_\al(h)=0,
\eq
\beq{2.12}
\left[r^2\frac{\p^2}{\p r^2}+3r\frac\p{\p r}+
r^2(\nu^2-1)-\al^2+1\right]\chi_{\nu,\al}(r)=0.
\eq
The solutions of (\ref{2.11}) and (\ref{2.12}) have the following form
$$
\chi_{\nu,\al}(r)=r^{-1}\clZ_{\al}(r\sqrt{\nu^2-1})\,,
$$
$$
v_\al(h)=
\left\{
\begin{array}{lc}
h^{-1}\clZ_{\al}(2i\ve^\oh|\mu|h^{-1}) & \ve=\pm 1\,,\\
h^{\al-1} & \ve=0\,.
\end{array}
\right.
$$
In this way we come to (\ref{elhar}), (\ref{elhar1}).
$\Box$

\bigskip
It follows from (\ref{2.7}) that the restrictions of the Klein-Gordon
equation
to the homogeneous spaces assume the form
\beq{lb}
\left(h^2\frac{\p^2}{\p h^2}+3\frac\p{\p h}+
4\ve h^{-2}\frac{\p^2}{\p\bz\p
z}\right)f_\nu(h,z,\bz)=(\nu^2-1)f_\nu(h,z,\bz)\,,
\eq
$$
H_3\to\ve=1\,,~~dS_3\to\ve=-1\,.
$$
Thus, we come to the following statement
\begin{cor}
The basic harmonics on H$_3\,(\bfL)$, dS$_3\,(\bfIL)$ and the light-cone
$\bfC^{1,3}$ are
\beq{ef}
f_\nu(\bz,h,z)=h^{-1}\exp(i\mu z+i\bmu\bz)
\clZ_{\nu^2-1}(2i\ve^\oh|\mu|h^{-1})\,,\ve=\pm 1\,,
\eq
and
\beq{ef1}
f_\nu(\bz,h,z)=h^{\al-1}\exp(i\mu
z+i\bmu\bz)\,,~~\ve=0\,,~~\al^2=\nu^2+1\,.
\eq
\end{cor}

\subsection{Scalar fields on $\mM^{2,2}$ in the horospheric coordinates}

It is easy to pass from the eigen-functions of the Klein-Gordon equation on
$\mM^{1,3}$ to the eigen-functions of the Klein-Gordon equation on
$\mM^{2,2}$. We come to the equation
\beq{2.100}
r^{-2}\left[h^2\frac{\p^2}{\p h^2}+3\frac\p{\p h}+
4\ve h^{-2}\frac{\p^2}{\p z_1\p z_2}-r^2\frac{\p^2}{\p r^2}-
3r\frac\p{\p r}\right] f_\nu(z_2,h,z_1;r)=\nu^2f_\nu(z_2,h,z_1;r)\,.
\eq

The analog of Proposition 3.1 has the form
\begin{predl}
The basic harmonics of the eigen-value problem (\ref{2.100}) are
\beq{elhar0}
f_\nu(\bz,h,z;r)=r^{-1}h^{-1}\exp(\mu_1 z_1+i\mu_2z_2)
\clZ_{\al}(r\nu)\clZ_{\al}(2i\ve^\oh\sqrt{\mu_1\mu_2}h^{-1})\,,\ve=\pm 1\,,
\eq
and
\beq{elhar10}
f_\nu(\bz,h,z;r)=h^{\nu-2}\exp(i\mu_1 z_1+i\mu_2z_2)\,,~~\ve=0\,,
~~\al^2=\nu^2+1\,,
\eq
where $\mu_1\mu_2,\al\in\mR$.
\end{predl}
Thus, we come to the scalar field on AdS$_3$ and on $\bfC^{2,2}$.
\begin{cor}
The basic harmonics on AdS$_3$ and the light-cone
$\bfC^{2,2}$ are
\beq{ef2}
f_\nu(\bz,h,z)=h^{-1}\exp(i\mu_1 z_1+i\mu_2z_2)
\clZ_{\nu^2-1}(2i\ve^\oh\sqrt{\mu_1\mu_2}h^{-1})\,,\ve=\pm 1\,,
\eq
and
\beq{ef12}
f_\nu(\bz,h,z)=h^{\al-1}\exp(i\mu_1
z_1+i\mu_2z_2)\,,~~\al^2=\nu^2+1\,,~~\ve=0\,.
\eq
\end{cor}


\section{ Non-commutative 4d Minkowski space $\mM^{1,3}_{\de,q}$.}
\setcounter{equation}{0}

\subsection{Definition}

We define an algebra generated by matrix elements of (\ref{0.1}).
\begin{defi}
The non-commutative 4d Minkowski space $\mM^{1,3}_{\de,q}$, $0<q\leq 1$
$\de\in\mN$
is the unital associative algebra with the anti-involution $*$ and
four generators $X_j,~j=1,\ldots,4$ with the quadratic relations
\beq{11.1}
X_1X_3=q^{-\de}X_3X_1\,,~~X_1X_2=q^\de X_2X_1\,,
\eq
\beq{12.1}
[X_2,X_3]=q^{\de-2}(1-q^2)X_1X_4\,,
\eq
\beq{13.1}
X_2X_4= q^{\de-2}X_4X_2\,,
\eq
\beq{14.1}
X_3X_4=q^{-\de+2}X_4X_3\,,
\eq
\beq{15.1}
[X_1,X_4]=0\,,
\eq
such that
\beq{16.1}
X_1^*=X_1,~~X_2^*=X_3,~~X_4^*=X_4\,.
\eq
\end{defi}

This space was described in \cite{AR}. Following this approach
we cast
the relations
in $\mM^{1,3}_{\de,q}$ in the form of
the reflection equation. Consider the basis in Mat$(2)$
$$
E_1=\mat{1}{0}{0}{0}\,,~E_2=\mat{
0}{1}{0}{0}\,,~E_3=\mat{0}
{0}{1}{0}\,,
~E_4=\mat{0}{0}{0}{1}\,.
$$
Define two R-matrices
\beq{R}
R(q)=q^{-1}(E_1\otimes E_1+E_4\otimes E_4)+( E_1\otimes E_4+E_4\otimes E_1)
+q^{-1}(1-q^2)E_3\otimes E_2\,,
\eq
\beq{R2}
R^{(2)}(q)=(E_1\otimes E_1+E_4\otimes E_4)+q^{\de-1}( E_1\otimes
E_4+E_4\otimes
E_1)\,.
\eq
The R-matrices satisfy the Yang-Baxter type equations
\beq{YB1}
R_{12}R_{13}R_{23}=R_{23}R_{
13}R_{12}\,,
\eq
\beq{YB2}
R_{12}R^{(2)}_{13}R^{(2)}_{23}=R^{(2)}_{
23}R^{(2)}_{13}R_{12}\,.
\eq
It can be checked straightforwardly that the relations
(\ref{11.1})-(\ref{15.1}) are equivalent to the reflection equation
\beq{RE}
R(q)\bfX^{(1)}R^{(2)}(q)\bfX^{(2)}=\bfX^{(2)}R^{(2)}(q)\bfX^{(1)}R^\dag(q)\,,
\eq
where $\bfX^{(1)}=\bfX\otimes Id$ and $\bfX^{(2)}=Id\otimes \bfX$ and
$$
R^\dag(q)=q^{-1}(E_1\otimes E_1+E_4\otimes E_4)+( E_1\otimes
E_4+E_4\otimes E_1)
+q^{-1}(1-q^2)E_2\otimes E_3\,.
$$
\begin{lem}
The algebra ${\mM}^{1,3}_{\de,q}$ has two independent Casimir elements
\beq{10.1}
K_1=X_1^{\de-2}X_4^\de\,,
\eq
\beq{10a.1}
K_2=X_1X_4-q^{-\de}X_3X_2\,.
\eq
\end{lem}
{\sl Proof.}\\
It can checked in straightforward way that the both expressions commute
with
the generators of ${\mM}^{1,3}_{\de,q}$.$\Box$

The Casimir operator $K_2$ (\ref{10.1}) is the quantum determinant
$K_2=\det_q\bfX$.
In an irreducible module over ${\mM}^{1,3}_{\de,q}$
$K_2=\ve r^2\in\mR$ is a
scalar. It allows us to define the
time-like part $\mM^{1,3+}_{\de,q}, ~(\ve=1)$, the space-like part
$\mM^{4-}_{\de,q}, ~(\ve=-1)$,
and the light cone $\bfC_{\de,q},~(\ve=0)$.

\subsection{Standard basis}

Represent the matrix $\bfX$ in the basis of the sigma-matrices
$$
\bfX=Y_0Id+\sum_{\al=}^3Y_\al\sigma_\al\,.
$$
The advantage of this basis is that its generators are self-conjugate with
respect to the anti-involution $Y_a^*=Y_a$.
Let $\te=\ln q$, $0<q\le1$ and
$$
c_1(\te,\de)=\oh\left(\cosh\te\de+\cosh\te(2-\de)\right)\,,~~
c_2(\te,\de)=\oh\left(\cosh\te\de-\cosh\te(2-\de)\right)\,,
$$
$$
c_3(\te\,\de)=\oh\left(\sinh\te\de+\sinh\te(2-\de)\right)\,,~~
c_4(\te,\de)=\oh\left(\sinh\te\de-\sinh\te(2-\de)\right)\,.
$$
In terms of $Y_a$ the
commutation relations in the algebra ${\mM}^{1,3}_{\de,q}$ assume the form
\beq{30.1}
Y_1Y_0=
c_1Y_0Y_1+c_2Y_3Y_1-ic_3Y_0Y_2-ic_4Y_3Y_2\,,
\eq
\beq{31.1}
Y_1Y_3=c_2Y_0Y_1+c_1Y_3Y_1-ic_4Y_0Y_2-ic_3Y_3Y_2\,,
\eq
\beq{32.1}
Y_2Y_0=ic_3Y_0Y_1+ic_4Y_3Y_1+c_1Y_0Y_2+c_2Y_3Y_2\,,
\eq
\beq{33.1}
Y_2Y_3=ic_4Y_0Y_1+ic_3Y_3Y_1+c_2Y_0Y_2+c_1Y_3Y_2\,,
\eq
\beq{34.1}
[Y_1,Y_2]=\frac{i}{2}q^{\de-2}(1-q^2)(Y_0^2-Y_3^2)\,,
\eq
\beq{35.1}
[Y_0,Y_3]=0\,.
\eq
Note that for $q\to 1$ $c_1(\te,\de)\to 1$ while $c_j\to 0$, $j=2,3,4$
and we come to the commutative space $\mM^{1,3}=(y_0,y_1,y_2,y_3)$.

The Casimirs assume the forms
$$
K_1=(Y_0+Y_3)^{\de-2}(Y_0-Y_3)^\de\,,
$$
$$
K_2=\left(1-\frac{1-q^{-2}}{2}\right)Y_0^2-
q^{-\de}Y_1^2-q^{-\de}Y_2^2-
\left(1-\frac{1-q^{-2}}{2}\right)Y_3^2\,.
$$

\subsection{Quantum Lorentz group action on $\mM^{1,3}_{\de,q}$.}

We start with a pair of the standard ${\cal U}_q({\rm SL}_2)$ Hopf algebra
\cite{KR,Sck}. The first one is generated by $A,B,C,D$ and the unit $1$
with relations
$$
AD=DA=1,~AB=qBA,~BD=qDB,
$$
\beq{3.1}
AC=q^{-1}CA,~CD=q^{-1}DC,
\eq
$$
[B,C]=\frac{1}{q-q^{-1}}(A^2-D^2).
$$

There is a copy of this algebra ${\cal U}^*_q({\rm SL}_2)$ generated by
$A^*,B^*,C^*,D^*$
with the relations coming from (\ref{3.1}) $U^*V^*=(VU)^*$. They commute
with $A,B,C,D$.

The pair ${\cal U}_q({\rm SL}_2)$, ${\cal U}^*_q({\rm SL}_2)$ forms
a Hopf algebra ${\cal U}^{(s)}_q({\rm SL}_2)$, where the coproduct and the
antipode are twisted in the
consistent way \cite{OR1}
$$
\D (A)=\D(A)=A\ot A,
$$
\beq{3.4}
\D (B)=A\otimes B+B\otimes D(A^*)^s,
\eq
$$
\D (C)=A\otimes C+C\otimes D(A^*)^{-s}.
$$
\beq{3.4a}
S \mat{A}{B}{C}{D}=
\mat{D}{-q^{-1}(A^*)^{-s}B}{-q(A^*)^sC}{A}.
\eq
$$
\D (A^*)=\D(A^*)=A^*\ot A^*,
$$
\beq{3.5}
\D (B^*)=A^*\otimes B^*+B^*\otimes D^*A^s,
\eq
$$
\D (C^*)=A^*\otimes C^*+C^*\otimes D^*A^{-s}.
$$
$$
S \mat{A^*}{B^*}{C^*}{D^*}=
\mat{D^*}{-q A^{-s}B^*}{-q^{-1}A^{s}C^*}{A^*}.
$$

The counit on $U^{(s)}_q(\SL)$ assumes the form
\beq{anti}
\ve (A)=1\,,~~\ve(B,C)=0\,.
\eq

There are two Casimir elements in $U^{(s)}_q(\SL)$ which commute with
any $u\in U_q(SL_2(\bf {C}))$.
\beq{3.6}
\Omega_q:=\frac{(q^{-1}+q)(A^2+A^{-2})-4}{2(q^{-1}-q)^2}+
\frac{1}{2}(BC+CB)
\eq
\beq{3.7}
\ti{\Omega}_q:=\frac{(q^{-1}+q)(A^{*2}+A^{*-2})-4}{2(q^{-1}-q)^2}+
\frac{1}{2}(B^*C^*+C^*B^*)
\eq

\begin{predl}
$\mM^{1,3}_{\de,q}$ is a right module over the Hopf algebra
${\cal U}^{(s)}_q(\SL)$.
\end{predl}
{\sl Proof.}\\
We define the action of the quantum group ${\cal U}^{(s)}_q(\SL)$
on ${\mM}^{1,3}_{\de,q}$
\beq{20.1}
\mat{X_1}{X_2}{X_3}{X_4}.A=\mat{q^\oh
X_1}{q^{-\oh}X_2}{q^\oh
X_3}{q^{-\oh}X_4}\,,
\eq
\beq{21.1}
\mat{X_1}{X_2}{X_3}{X_4}.B=\mat{
0}{X_1}{0}{X_3}\,,
\eq
\beq{22.1}
\mat{X_1}{X_2}{X_3}{X_4}.C=\mat{
X_2}{0}{X_4}{0}\,,
\eq
\beq{23.1}
\mat{X_1}{X_2}{X_3}{X_4}.A^*=
\mat{q^{\frac{1-\de}{s}} X_1}{q^{\frac{1-\de}{s}}
X_2}{q^{\frac{\de-1}{s}}X_3}{q^{\frac{\de-1}{s}}X_4}\,.
\eq
The direct calculations show that the commutation relations in
$\mM^{1,3}_{\de,q}$
are compatible with the coproduct in ${\cal U}^{(s)}_q(\SL)$.
Moreover,
$$
(X_j.a)^*=a^*.X_j^*\,.
$$
$\Box$

\bigskip

Similarly, one can define the left action of ${\cal U}^{(s)}_q(\SL)$
on $\mM^{1,3}_{\de,q}$.

Let
$$
w(m,k,l,n)=X_3^mX_1^kX_4^lX_2^n
$$
be the ordered monomial.
Define the Schwartz space $\clS(\mM^{1,3}_{\de,q})$ as the series
with the rapidly decreasing coefficients
\beq{Sch}
\clS(\mM^{1,3})=\{\ddag
f(X_3,X_1,X_4,X_2)\ddag=\sum_{m,k,l,n}a_{m,k,l,n}w(m,k,l,n)\,,~
a_{m,k,l,n} \in\mC\}\,,
\eq
$$
|a_{m,k,l,n}|<(1+m^2+k^2+l^2+n^2)^j,~{\rm for~any~}j\in\mN,~{\rm when}~
|m|,|k|,|l|,|n|\to\infty\,.
$$
\begin{predl}
The Jackson integral
\beq{ji}
\lan f \ran=\int
d_{q^2}X_3d_{q^2}X_1d_{q^2}X_4d_{q^2}X_2\ddag f(X_1,X_2,X_3,X_4)\ddag
\eq
is invariant functional on $\clS(\mM^{1,3}_{\de,q})$ with respect to the
action of
${\cal U}^{(s)}_q(\SL)$
$$
\lan f.u \ran=\ve(u)\lan f \ran \,,
$$
where $\ve(u)$ is the counit (\ref{anti}).
\end{predl}
The proof will follow from the actions of the generators $A,A^*,B,C$ on the
ordered monomials $w(m,k,l,n)$ presented in Section 9.


\section{Module over ${\mM}^{1,3}_{\de,q}$}
\setcounter{equation}{0}

Here we construct representations of ${\mM}^{1,3}_{\de,q}$ in the
infinite-dimensional spaces $E^+,E^-,E^0$ (right modules over
${\mM}^{1,3}_{\de,q}$). The representations of algebras satisfying the more
general
reflection equations were constructed in Ref.\,\cite{K}.
Three types of the spaces correspond to
the time-like part $K_2>0$, the space-like part $K_2<0$, and the light
cone $K_2=0$ .
Here we assume that $\de\in\mZ$.
\bigskip

\subsection{Time-like part ($K_2>0$).}

We will define the module $E_{\al,\rho}^+$, depending on $\al,\rho\in\mR$.
Consider $\mC$-valued functions on the $\mZ$ lattice with the vertices
$q^j$.
Using the Jackson integral we introduce the Hermitian metric
\beq{1.2}
\langle f|g\rangle=\int
d_{q^2}ud_{q^2}\bu\bfe_{q^2}(u\bu)\overline{f(u)}g(u)\,,
\eq
where $\bfe_{q^2}(u)=e_{q^2}((1-q^2)u)$ (\ref{exp}) and the Jackson
integral (\ref{J}) is defined by the double series
\beq{1.2a}
\langle
f|g\rangle=(1-q^2)^2\sum_{m,n\in\mZ}q^{2(m+n)}[\bfe_{q^2}(q^{2(m+n)})\overline{f(q^{
2m})}g(q^{2n})
+\bfe_{q^2}(-q^{2(m+n)})\overline{f(-q^{2m})}g(-q^{2n})]\,.
\eq

The module $E_{\al,\rho}^+$ is the
space of functions on the lattice with finite norm $\langle
f|f\rangle^\oh<\infty$.
Let $\la=\la(\de,q)=q^{\frac{\de}{2}-1}(1-q^2)^\oh$.
\begin{predl}
The space $E_{\al,\rho}^+$ ($\rho\in\mR^+,\,\al\in\mR^+$) is the right
module
of the algebra ${\mM}^{1,3}_{\de,q}$
\beq{4.2}
f(u)X_1=\al^{-1}\rho f(uq^\de)\,,~~~f(u)X_2=\la\rho D_{u}f(u)\,,
\eq
$$
f(u)X_3=-\la\rho uf(u)\,,~~~f(u)X_4=\al\rho f(uq^{2-\de})\,.
$$
The Casimirs on $ E_{\al,\rho}^+$ have the values $K_2=\rho^2q^{-2}=R^2$,
$K_1=\al^2\rho^{2\de-2}$.
The anti-involution corresponds to the Hermitian conjugation
\beq{5.2}
\langle f X_a|g\rangle=\langle f|X_a^*g\rangle\,.
\eq
$D_u$ is defined by (\ref{dif}).
\end{predl}
{\sl Proof}\\
We have the following relations for $T_u:~f(u)\to f(uq)\,,~~~u:~f(u)\to
uf(u)$
and the for difference operator
(\ref{dif})
\beq{2.2}
f(u)(u\cdot T_q)=qf(u)(T_q\cdot u)\,,
~~f(u)(D_{u}\cdot T_q)=q^{-1}f(u)(T_q\cdot
D_u\,,
\eq
$$
f(u)[D_{u},u]=-f(uq^2)\,.
$$
They imply the commutation relations (\ref{11.1}) - (\ref{15.1}) for
(\ref{4.2}).
The conjugation formula $X_2^*=X_3$ comes from the relation
$$
D_{u}\bfe_{q^2}(au)=a\bfe_{q^2}(au)\,.
$$
The shift operators $X_1$ and $X_4$ are self-conjugate due to (\ref{1.2a}).
$\Box$

Note that the operator $X_1+X_4$ acts as
$$
f(u)\to \rho(\al^{-1}f(uq^\de)+\al(uq^{2-\de}))\,.
$$
Since $\al$ and $\rho$ are positive this operator is positive definite
$$
\langle (X_1+X_4)f|f\rangle>0\,,~~{\rm if}~f\neq 0\,.
$$
This fact is in agreement with the classical limit $\oh(X_1+X_4)\to
y_0$, and
$y_0>0$.

\subsection{Space-like part ($K_2<0$).}

Replace the measure in the integral (\ref{1.2})
\beq{7.2}
\langle f|g\rangle=\int
d_{q^2}ud_{q^2}\bu\bfe_{q^2}(-u\bu)\overline{f(u)}g(u)\,.
\eq
It defines the Hermitian pairing in the module $E_{\al,\rho}^-$.
\begin{predl}
The space $E_{\al,\rho}^-$, $(\al,\rho\in\mR^+)$ is the right module of
the algebra ${\mM}^{1,3}_{\de,q}$
\beq{8.2}
f(u)X_1=\al^{-1}\rho f(uq^\de)\,,~~~f(u)X_2=\la\rho D_{u}f(u)\,,
\eq
$$
f(u)X_3=\la\rho uf(u)\,,~~~f(u)X_4=-\al\rho f(uq^{2-\de})\,.
$$
with $K_2=-R^2=-\rho^2q^{-2}$ and $K_1=(-1)^\de\al^2\rho^{2\de-2}$.
\end{predl}
The proof of this Proposition is the same as the previous one. We just
change consistently the sign in the measure and in the definition of
$X_3$, $X_4$.

In this case the operator $X_1+X_4$ is not positive-definite in accordance
with the classical description of the space-like part.

\subsection{The light-cone $K_2=0$.}

Consider the space $E^0$ of holomorphic functions on $\mC^*$
equipped with the Hermitian form
\beq{9.2}
\langle f|g\rangle=\f1{2\pi i}\oint \frac{dw}{w}\overline{f(w)}g(w)\,.
\eq
The following Proposition can be established in the direct way.
\begin{predl}
$E^0$ is the right module over the light-cone $\bfC^{1,3}_{\de,q}$
\beq{10.2}
f(w)X_1=f(wq^\de)\,,~~~f(w)X_2=q^\frac{\de-1}{2}w^{-1}f(wq)\,,
\eq
$$
f(w)X_3=q^\frac{\de-1}{2}wf(wq) \,,~~
f(w)X_4=f(wq^{2-\de})\,,
$$
with $K_2=0$ and $K_1=1$.
\end{predl}



\section{Horospheric description.}
\setcounter{equation}{0}
\subsection{Horospheric generators}

We introduce another set of generators - the non-commutative analog of
the horospheric
coordinates
$(Z^*,H,Z,R),~(H^*=H,\,R^*=R,\,(Z^*)^*=Z)$
\beq{5.1}
X_1=RH\,,~~X_2=RHZ\,,~~X_3=RZ^*H\,,
\eq
\beq{5a.1}
X_4=R(Z^*HZ+\ve H^{-1})\,,~~\ve=\pm1, ~0\,.
\eq
The defining relations
\beq{hcr}
ZH=q^{-\de} HZ\,,~~Z^*H=q^{\de}HZ^*\,,~~[R,H]=[R,Z]=[R,Z^*]=0\,,
\eq
$$
ZZ^*=q^{2\de-2}Z^*Z-\ve q^{\de-2}(1-q^2)H^{-2}\,.
$$
yield the relations (\ref{11.1})--(\ref{15.1}).
The Casimir elements are
\beq{Ca}
K_2=\ve R^2\,~~~K_1=R^{2\de-2}H^{\de-2}(Z^*HZ+\ve H^{-1})^\de\,.
\eq

The inverse relations assume the form
\beq{40.1}
H=R^{-1}X_1\,,~~~Z=X_1^{-1}X_2\,,~~~Z^*=X_3X_1^{-1}\,,~~R=\ve K_1\,.
\eq

\bigskip
In terms of the horospheric generators the action of ${\cal
U}^{(s)}_q(\SL)$ takes the
form
\beq{3.11}
\begin{array}{llll}
Z^*.A=z^*\,, & H.A=q^{\oh}H\,, & Z.A=q^{-1}z\,, & R.A=R\,,\\
Z^*.A^*=q^{\frac{2\de-2}{s}}Z^*\,, &
H.A^*=q^{\frac{\de-1}{s}} H\,, & Z.A^*=Z\,, & R.A^*=R\,,\\
Z^*.B=0\,, & H.B=0\,, & Z.B=q^{-\oh}\,, & R.B=0\,, \\
Z^*.C=q^{\frac{3}{2}-\de}H^{-2}\,, & H.C=HZ\,, & Z.
C=-q^{\oh}Z^2\,,&
R.C=0\,.
\end{array}
\eq
It follows from these relations that $R$ is invariant with respect to
the ${\cal U}^{(s)}_q(\SL)$
action $R.u=\ve(u)R$.

Define the analog of
the Schwartz space $\clS(\mM^{1,3}_{\de,q})$ (\ref{Sch}) in
terms of the ordered monomial $\hat{w}(m,k,n)=Z^{*m}H^kZ^n$. Since $R$
is a center
element its
position is irrelevant. Let
\beq{Sch1}
\ddag f(Z^*,H,Z,R)\ddag=\sum_{m,k,n,l}a_{m,k,n,l}\hat{w}(m,k,n)R^l\,,~~
a_{m,k,n,l} \in\mC\,.
\eq
For $\clS(\mM^{1,3}_{\de,q})$ the coefficients satisfy the condition
$$
|a_{m,k,n,l}|<(1+m^2+k^2+l^2+n^2)^j,~{\rm for~any~}j\in\mN,~{\rm when}~
|m|,|k|,|l|,|n|\to\infty\,.
$$
The invariant integral (\ref{ji}) is well defined functional on
(\ref{Sch1}). It assumes the form
\beq{intq}
I_{q^2}(f)=\int d_{q^2}Z^*d_{q^2}Hd_{q^2}Zd_{q^2}R\ddag
f(Z^*,H,Z,R)H\ddag\,.
\eq

\subsection{Homogeneous spaces}

Consider an irreducible representation of algebra (\ref{hcr}).
Then one can fix the Casimir operator (\ref{Ca}) $K_2=\ve R^2$,
$R^2=r^2\in\mR^+$.
It allows us to define the non-commutative analog of Lobachevsky spaces
and the
cone. Let us fix the ideal $I_\ve=\{K_2-\ve r^2=0\}$. Then
$$
\clS(\mM^{1,3}_{\de,q})/I_\ve\sim~ H_3\,(\ve=1)\,,~dS_3\,(\ve=-1)\,,~{\bf
C}^{1,3}_{\de,q}\,(\ve=0)\,.
$$

As we observed above the action
of the quantum Lorentz group preserves these spaces. It justifies
the notion of homogeneous spaces in the noncommutative situation.

We can directly define their generators using the horospheric
description of
$\mM^{1,3}_{\de,q}$.
\begin{defi}
The non-commutative Lobachevsky space $\bfI_{\de,q}$ (H$_3$),
the non-commutative Imaginary Lobachevsky space $\bf{IL}_{\de,q}$
(dS$_3$) and the non-commutative cone ${\bf C}^{1,3}_{q,\de}$
are
the associative unital
algebras with an anti-involution and the defining relations
$$
ZH=q^{-\de} HZ\,,~~Z^*H=q^{\de}HZ^*\,,
$$
$$
ZZ^*=q^{2\de-2}Z^*Z-\ve q^{\de-2}(1-q^2)H^{-2}\,.
$$
$$
(Z)^*=Z^*\,,~~H^*=H\,,
$$
$$
{\rm H}_3\,\sim\,\ve= 1\,,~~{\rm dS}_3\,\sim\,\ve= -1\,,~~
\bfC^{1,3}_{\de,q}\,\sim\,\ve= 0\,.
$$
\end{defi}

In addition we define the non-commutative absolute.
\begin{defi}
The non-commutative absolute ${\bf\Xi}_{\de,q}$
is the associative algebra with two generators and the commutation
relation
\beq{abs}
ZZ^*=q^{-2+2\de}Z^*Z\,.
\eq
\end{defi}

\subsection{Representations of the horospheric generators}

For the time-like part $\ve=1$ we considered the
space $E_{\al,\rho}^+$. We have
from (\ref{4.2}), (\ref{5.1}), (\ref{5a.1})
\beq{6.2}
f(u)H=\rho f(uq^\de)\,,
\eq
$$
f(u)Z=\al^{-1}\la\frac{f(uq^{-\de})-f(uq^{2-\de})}
{(1-q^2)}u^{-1}\,,~~~
f(u)Z^*=-\al^{-1}q^{-\de}\la uf(uq^{-\de})\,,
$$
and $K_2=\rho^2q^{-2}$.

Similarly, the representation of the space-like part $(\ve=-1)$
of $\mM^{1,3}_{\de,q}$ in the space $E_{\al,\rho}^-$
assumes the the following form.
The horospheric generators $H,Z$ are represented as before (\ref{6.2}),
while
$$
f(u)Z^*=q^{-\de}\la uf(uq^{-\de})\,.
$$

The horospheric generators corresponding to the light cone
$\bfC^{1,3}_{\de,q}$
act in the space $E_0$ (\ref{9.2}) as
\beq{10.3}
f(w)H=f(wq^\de)\,,
\eq
\beq{10.4}
f(w)Z= q^{\frac{\de-1}{2}}wf(wq^{1-\de})\,,
~~~f(w)Z^*=q^\frac{3\de-1}{2}w^{-1}f(wq^{1-\de})\,.
\eq

The module $E^0$ serves simultaneously as the module of the non-commutative
absolute ${\bf\Xi}_{\de,q}$. It is easy to derive from
(\ref{10.4}) that the generators $Z,Z^*$
that satisfy (\ref{abs}) are represented as follows
\beq{10.5}
f(w)Z=wf(wq^{1-\de})\,,
~~~f(w)Z^*=w^{-1}f(wq^{1-\de})\,.
\eq

\section{ Non-commutative Minkowski space $\mM^{2,2}_{q}$.}
\setcounter{equation}{0}

\subsection{Definition}
The space $\mM^{2,2}_{q}$ is related to the Hopf algebra
${\cal U}_q({\rm SL}_2(\mR))$. It implies that $|q|=1$, ($q=\exp i\te$).
In this case the quadratic relations (\ref{11.1}) -- (\ref{15.1}) are
valid for $\de=1$
only.

\begin{defi}
The non-commutative 4d Minkowski space $\mM^{2,2}_{q}$, $(|q|=1)$
is the unital associative *-algebra with
four generators $X_j,~j=1,\ldots,4$ that satisfies the quadratic relations
\beq{51.1}
X_1X_3=q^{-1}X_3X_1\,,~~X_1X_2=q X_2X_1\,,
\eq
\beq{52.1}
[X_2,X_3]=(q^{-1}-q)X_1X_4\,,
\eq
\beq{53.1}
X_2X_4= qX_4X_2\,,
\eq
\beq{54.1}
X_3X_4=q^{-1}X_4X_3\,,
\eq
\beq{55.1}
[X_1,X_4]=0\,,
\eq
and with the anti-involution
\beq{inv}
X_1^*=X_1\,,~X_2^*=q^{-1}X_2\,,~X_3^*=qX_3\,,~X_4^*=X_4\,.
\eq
\end{defi}

The reflection equation describing these relations is reduced to the
standard RTT form for the Hopf algebra ${\cal A}_q({\rm GL}_2(\mR))$
\beq{RTT}
R(q)\bfX^{(1)}\bfX^{(2)}=\bfX^{(2)}\bfX^{(1)}R^\dag(q)\,,
\eq
since $R^{(2)}=Id\otimes Id$ for $\de=1$ and
$$
R(q)=q^{-1}(E_1\otimes E_1+E_4\otimes E_4)+( E_1\otimes E_4+E_4\otimes E_1)
+q^{-1}(1-q^2)E_3\otimes E_2\,.
$$
In the classical limit ${\cal A}_q({\rm GL}(2,\mR))$ passes
to the algebra ${\cal A}({\rm GL}_2(\mR))$. It allows
to identify $\mM^{2,2}_{q}$ with a non-commutative deformation of the
Minkowski space $\mM^{2,2}$.

 From Lemma 4.1 we have the form of the Casimir elements of $\mM^{2,2}_{q}$
\beq{60.1}
K_1=X_1^{-1}X_4\,,
\eq
\beq{60a.1}
K_2=X_1X_4-q^{-1}X_3X_2\,.
\eq
Note that $K_2$ is not self adjoint. The self adjoint element is
$\ti{K}_2=(1+q^2)K_2$.

\subsection{Quantum group ${\cal U}_q({\rm SL}_2(\mR))\oplus{\cal U}_q({\rm
SL}_2(\mR))$
action on $\mM^{2,2}_{q}$.}

Consider the Hopf algebra ${\cal U}_q({\rm SL}_2)$ (\ref{3.1}).
The following conditions pick up the *-Hopf algebra
${\cal U}_q({\rm SL}_2(\mR))$
$$
A^*=A\,,~~B^*=-B\,,~~C^*=-C
$$
with the untwisted coproduct (\ref{3.4})
$$
\D (A)=\D(A)=A\ot A,
$$
$$
\D (B)=A\otimes B+B\otimes D\,,
$$
$$
\D (C)=A\otimes C+C\otimes D\,,
$$
and the untwisted antipode (\ref{3.4a})
\beq{ant}
S \mat{A}{B}{C}{D}=
\mat{D}{-q^{-1}B}{-qC}{A}\,.
\eq

The Casimir element of ${\cal U}_q({\rm SL}_2(\mR))$
has the form
\beq{3.6a}
\Omega_q:=\frac{(q^{-1}+q)(A^2+A^{-2})-4}{2(q^{-1}-q)^2}+
\frac{1}{2}(BC+CB)
\eq

\begin{predl}
$\mM^{2,2}_{\de,q}$ is the module over the *-Hopf algebra
${\cal U}_q({\rm SL}_2(\mR))$.
\end{predl}
{\sl Proof}\\
Define the right action of ${\cal U}_q({\rm SL}_2(\mR))$ on $\mM^{2,2}_{q}$
\beq{70.1}
\mat{X_1}{X_2}{X_3}{X_4}.A=\mat{q^\oh
X_1}{q^{-\oh}X_2}{q^\oh
X_3}{q^{-\oh}X_4}\,,
\eq
\beq{71.1}
\mat{X_1}{X_2}{X_3}{X_4}.B=\mat{
0}{X_1}{0}{X_3}\,,
\eq
\beq{72.1}
\mat{X_1}{X_2}{X_3}{X_4}.C=\mat{
X_2}{0}{X_4}{0}\,.
\eq

The consistency of multiplication in $\mM^{2,2}_{\de,q}$ and the
comultiplication in ${\cal U}_q({\rm SL}_2(\mR))$ follows from Proposition
4.1, since (\ref{70.1}) -- (\ref{72.1}) comes from (\ref{20.1}) --
(\ref{22.1}).

We also define the left action of
${\cal U}_q({\rm SL}_2(\mR))$ on $\mM^{2,2}_{q}$
$$
A.\mat{X_1}{X_2}{X_3}{X_4}=\mat{q^{-\oh}X_1}{q^{-\oh}X_2}{q^\oh X_3}{q^\oh
X_4}\,,
$$
$$
B.\mat{X_1}{X_2}{X_3}{X_4}=\mat{-X_3}{-X_4}{0}{0}\,,
$$
$$
C.\mat{X_1}{X_2}{X_3}{X_4}=\mat{0}{0}{-X_1}{-X_2}\,.
$$
It is also consistent with the multiplication in $\mM^{2,2}_{q}$.

In addition from (\ref{inv}) and (\ref{ant}) one has
$$
S(u)^*.X_j^*=(X_j.u)^*\,,~~u=A\,,\,B\,,\,C\,.
$$
$\Box$

\subsection{Horospheric generators and homogeneous spaces}

The non-commutative analog of the horospheric generators $(Z_2,H,Z_1,R)$
on $\mM^{2,2}_{q}$ assumes the form
\beq{5.1a}
X_1=RH\,,~~X_2=RHZ_1\,,~~X_3=RZ_2H\,,
\eq
\beq{5a.1a}
X_4=R(Z_2HZ_1+\ve H^{-1})\,,~~\ve=\pm1, ~0\,.
\eq
It follows from (\ref{inv}) that
$$
Z_2^*=Z_2\,,~H^*=H\,,~Z_1^*=Z_1\,,~R^*=R\,.
$$
The defining relations
\beq{hcr1}
Z_1H=q^{-1} HZ_1\,,~~Z_2H=qHZ_2\,,~~[R,H]=[R,Z_1]=[R,Z_2]=0\,,
\eq
$$
[Z_1,Z_2]=\ve (q-q^{-1})H^{-2}\,.
$$
yield the relations (\ref{51.1})--(\ref{55.1}).
The Casimir elements are
\beq{Ca1}
K_2=\ve R^2\,~~~K_1=H^{-1}(Z_2HZ_1+\ve H^{-1})\,.
\eq

The horospheric coordinates can be expressed in
terms of $X_j$ as
\beq{40.1a}
H=R^{-1}X_1\,,~~~Z_1=X_1^{-1}X_2\,,~~~Z_2=X_3X_1^{-1}\,,~~R^2=\ve K_2\,.
\eq

\bigskip
In terms of the horospheric generators the right action of
${\cal U}_q({\rm SL}_2(\mR))$ takes the
form
\beq{80.11}
\begin{array}{llll}
Z_2.A=Z_2\,, & H.A=q^{\oh}H\,, & Z_1.A=q^{-1}Z_1\,, & R.A=R\,,\\
Z_2.B=0\,, & H.B=0\,, & Z_1.B=q^{-\oh}\,, & R.B=0\,, \\
Z_2.C=q^{\frac{1}{2}}H^{-2}\,, & H.C=HZ_1\,, & Z_1.C=-q^{\oh}Z_1^2\,,&
R.C=0\,.
\end{array}
\eq
It follows from these relations that $R$ is invariant with respect to
the ${\cal U}_q({\rm SL}_2(\mR))$ action.

\bigskip
Let ${\cal S}(\mM^{2,2})$ be the algebra of the ordered Schwartz functions
on $\mM^{2,2}_{q}$.
Consider an irreducible representation of the associative algebra
$\mM^{2,2}_{q}$.
We fix the value of the Casimir operator (\ref{Ca1}) $K_2=\ve R^2$,
$R^2=r^2\in\mR^+$
and the ideal $I_\ve=\{K_2-\ve R^2=0\}$.
It allows us to define the non-commutative analog of AdS$^\pm_3$ and the
cone $\bfC^{2,2}_q$ as
$$
{\cal S}(\mM^{2,2})/I_\ve\,\sim\,~~ AdS_3^\pm\,(\ve=\pm
1)\,,~\bfC_q^{2,2}\,(\ve=0)\,.
$$
The direct description of these algebras in terms of the horospheric
generators has the form
\begin{defi}
The non-commutative AdS$_3^\pm$ spaces and the non-commutative cone
$\bfC^{2,2}_{q}$ are the associative unital
algebras with the anti-involution and the defining relations
$$
Z_1H=q^{-1} HZ_1\,,~~Z_2H=qHZ_2\,,
$$
$$
[Z_1,Z_2]= \ve (q-q^{-1})H^{-2}\,,
$$
where AdS$^\pm_3\to(\ve=\pm 1)$ and $\bfC^{2,2}_{q}\to
(\ve=0)$.
\end{defi}

\subsection{The non-commutative Minkowski space
$\widetilde{\mM}^{2,2}_{q}$}

\begin{defi}
The non-commutative 4d Minkowski space $\widetilde{\mM}^{2,2}_{q}$,
$q\in(0,1]$
is the unital associative *-algebra with
four generators $X_j,~j=1,\ldots,4$ that satisfies the quadratic relations
(\ref{51.1}) -- (\ref{55.1})
with the anti-involution
\beq{inv2}
X_1^*=X_4\,,~X_2^*=qX_3\,.
\eq
\end{defi}
This algebra was considered in Ref.\,\cite{V}.

The commutation relations are the same as for ${\mM}^{2,2}_{q}$ and the
only
difference is the anti-involution (compare (\ref{inv}) and (\ref{inv2})
and the reality of $q$.
Therefore, we have the same Casimirs $K_1$ (\ref{60.1}) and $K_2$
(\ref{60a.1}). It allows us to define the associative algebras
$$
\widetilde{\rm AdS}_3^\pm ~\to~ K_2=\pm r^2\neq 0\,,
$$
$$
\widetilde{\bf C}^{2,2}_q~\to~K_2=0\,.
$$

\bigskip

Define the Hopf algebra ${\cal U}_q(\SUo)$
by the anti-involution
\beq{inv3}
A^*=A\,,~~B^*=-C\,,~~C^*=-B\,.
\eq
The Hopf algebra ${\cal U}_q(\SUo)\oplus{\cal U}_q(\SUo) $ acts on
$\widetilde{\mM}^{2,2}_{q}$ in the similar way as
${\cal U}_q({\rm SL}_2(\mR))\oplus{\cal U}_q({\rm
SL}_2(\mR))$
acts on $\mM^{2,2}_{q}$ (Proposition 7.1). The anti-involutions in
${\cal U}_q(\SUo)$ and in
$\widetilde{\mM}^{2,2}_{q}$ are consistent $S(u)^*.X_j^*=(X_j.u)^*$,
$u\in{\cal
U}_q(\SUo)$. It can be checked that $K_2$ is invariant with respect to this
action. Therefore, the action of ${\cal U}_q(\SUo)\oplus{\cal U}_q(\SUo)
$ is well
defined on $\widetilde{\rm AdS}_3^\pm$ and $\widetilde{\bf C}_q^{2,2}$.
In the classical limit we come to $\mM^{2,2}$ with the signature
$(+,-,-,+)$.


\section{Quasi-classical description of the non-commutative Minkowski
spaces.}
\setcounter{equation}{0}

\subsection{Quadratic Poisson algebras}

Consider first the Minkowski space $\mM^{1,3}_{\de,q}$.
In the limit $q\to 1$ we come in the first order of $\te=\log q$ to the
Poisson
algebra ${\cal M}^{1,3}$. It is a commutative algebras of
functions on $\mM^{1,3} $
equipped with the quadratic Poisson brackets
$$
\{x_j,x_k\}_2=\lim_{\te\to 0}\frac{X_jX_k-X_kX_j}{\te}\,.
$$
Explicitly the brackets take the form
\beq{1.p}
\{x_1,x_2\}_2=\de x_1x_2\,,~~~\{x_1,x_3\}_2=-\de x_1x_3\,,
\eq
\beq{2.p}
\{x_4,x_2\}_2=(2-\de)x_2x_4\,,~~~\{x_4,x_3\}_2=-(2-\de)x_3x_4\,,
\eq
\beq{3.p}
\{x_2,x_3\}_2=-2x_1x_4\,,~~~\{x_1,x_4\}_2=0\,.
\eq
In addition, there is the anti-involution on ${\cal M}^{1,3}$ that comes
from the anti-involution on the algebra $\mM^{1,3}_{\de,q}$
\beq{cinv}
\{x_j,x_k\}_2^*=-\{x^*_j,x^*_k\}_2\,.
\eq
It takes the form
$$
x^*_1=x_1\,,~~ x^*_4=x_4\,, ~~x^*_2=x_3\,.
$$
There are two Casimir functions
\beq{4.p}
k_1=x_1x_4-x_2x_3\,,~~~k_2=x_1^{\de-2}x_4^\de\,.
\eq

The Poisson algebra ${\cal M}^{1,3}$ can be represented in the form of the
classical reflection equation. Let $\te=\ln q$.
The quantum $R$ matrices (\ref{R2}) have the expansion
$R=Id\otimes Id+\te r$, $R^{(2)}=Id\otimes Id+\te r^{(2)}$, where
\beq{5.p}
r=-(E_1\otimes E_1+E_4\otimes E_4)-2 E_3\otimes E_2\,,
\eq
and
\beq{6.p}
r^{(2)}= (\de-1)(E_1\otimes E_4+E_4\otimes E_1)
\eq
Then the reflection equation in the quasi-classical limit gives
\beq{7.p}
\{\bfx_1,\bfx_2\}_2=r(\bfx_1\otimes\bfx_2)-(\bfx_2\otimes\bfx_1)r^\dag
+\bfx_2r^{(2)}\bfx_1-\bfx_1r^{(2)}\bfx_2\,.
\eq
It can be checked that the Poisson structure (\ref{1.p})--(\ref{3.p})
can be extracted from this equation.

We rewrite the Poisson algebra ${\cal M}^{1,3}$ in terms of the horospheric
coordinates (\ref{2.0}), (\ref{2a.0})
\beq{8.p}
\{h,z\}_2=\de hz\,,~~~\{h,\bz\}_2=-\de
h\bz\,,~~~\{r,h\}_2=\{r,z\}_2=\{r,\bz\}_2=0\,,
\eq
\beq{9.p}
\{z,\bz\}_2=2(\de-1)z\bz-2\ve h^{-2}\,.
\eq
The Casimir functions are
$$
k_1=\ve r^2\,,~~~k_2=h^{-2}(h^2|z|+\ve)^\de\,.
$$
The symplectic leaves of this structure are two-dimensional surfaces
$$
\ve r^2=c_1\,,~~~h^{-2}(h^2|z|+\ve)^\de=c_2\,.
$$
If $\de=0,2$ the symplectic leaves are the horospheres,
i.e. the orbits of unipotent subgroups. For $\de=1$ the leaves are
the orbits of groups conjugated to SU(2) $(\ve=1)$ or SU(1,1) $(\ve=-1)$.

\bigskip
The quadratic Poisson algebra $\widetilde{\cal M}^{2,2}$ that arises in
the case
$\widetilde{\mM}^{2,2}_q$ coincides with (\ref{1.p}) -- (\ref{3.p}) for
$\de=1$.
It differs from ${\cal M}^{1,3}$ by the anti-involution
$$
x^*_1=x_4\,,~ ~~x^*_2=x_3\,.
$$

In the case of ${\mM}^{2,2}_q$ one should put $q=\exp i\te$. Therefore
we come
to the Poisson algebra ${\cal M}^{2,2}$
$$
\{x_1,x_2\}_2=i x_1x_2\,,~~~\{x_1,x_3\}_2=-i x_1x_3\,,
$$
$$
\{x_4,x_2\}_2=ix_2x_4\,,~~~\{x_4,x_3\}_2=-ix_3x_4\,,
$$
$$
\{x_2,x_3\}_2=-2ix_1x_4\,,~~~\{x_1,x_4\}_2=0\,.
$$
with the anti-involution $x_j^*=x_j$.
In the last two cases the classical reflection equations are simplified
since $r^{(2)}=Id\otimes Id$.

\subsection{Trihamiltonian structure}

In addition to the quadratic brackets we introduce the linear and the
canonical
brackets in the space of functions on the Minkowski spaces. Consider for
brevity the case ${\cal M}^{1,3}$. The linear brackets assume the form
\beq{1.lb}
\{x_1,x_2\}_1=\de x_2\,,~~~\{x_1,x_3\}_1=-\de x_3\,,
\eq
\beq{2.lb}
\{x_4,x_2\}_1=(\de-2)x_2\,,~~~\{x_4,x_3\}_1=(2-\de)x_3\,,
\eq
\beq{3.lb}
\{x_2,x_3\}_1=-2(x_1-x_4)\,,~~~\{x_1,x_4\}_1=0\,.
\eq
For $\de=1$ it is the Lie-Poisson brackets on the Lie algebra gl$(2,\mR)$.
Two Casimir functions of this structure are
$$
k_1=(\de-2)x_1-\de x_4\sim y_0-(\de-1)y_3\,,
$$
$$
k_2=x_2x_3-(x_1-x_4)^2\sim y_1^2-y_2^2-2y_3^2\,.
$$
The linear brackets (\ref{1.lb})-(\ref{3.lb}) can be reproduced in terms of
the canonical
variables $(v,u)$, $\{v,u\}=1$
$$
x_1=-\de uv-\oh k_1\,,~~x_2=v\,,~~x_3=2u^2v\,,
$$
$$
x_4=(2-\de)uv-\oh k_1\,.
$$
The quantization of the linear brackets is straightforward $u\to U$,
$v\to V=\p_U$,
where $V$ is standing on the right.

The canonical brackets are
\beq{1.cb}
\{x_2,x_3\}_0=-2\,,~~~\{x_1,x_j\}_0=\{x_4,x_j\}_0=0\,.
\eq

Note that the both Poisson structure are anti-invariant with respect to the
involution.

\begin{lem}
The Poisson structures on ${\cal M}^{1,3}$ are compatible.
\end{lem}
It means that their linear
combination
\beq{11.p}
\{~,~\}_\la=\{~,~\}_2+\la\{~,~\}_1+\la^2\{~,~\}_0\,~~\la\in\mC
\eq
is again a Poisson brackets. To prove it we
introduce the new algebra ${\cal M}^{1,3}_\la$ with the shifted generators
\beq{10.p}
x_1\to x_1+\la\,,~~x_4\to x_4-\la\,,~~x_2\to x_2\,~~x_3\to x_3\,.
\eq
This shift does not spoil the Jacobi identity of the brackets and we
come to
(\ref{11.p}). $\Box$


\section{The Laplace operator and its eigen-functions.}
\setcounter{equation}{0}

In this section we consider in details the case $\mM_{\de,q}^{1,3}$ and
shortly
reproduce the similar construction for $\mM_q^{2,2}$. The case
$\widetilde{\mM}_q^{2,2}$
will not be considered in this paper.

\subsection{The Laplace operator on $\mM_{\de,q}^{1,3}$}

Consider the action of ${\cal U}_q^{(s)}({\rm SL}(2,\mC))$ on the
ordered monomials $w(m,k,l,n)=X_2^mX_1^kX_4^lX_2^n$.
It follows from (\ref{20.1}) - (\ref{23.1}) that
$$
w(m,k,l,n).A=q^{\frac{m+k-l-n}2}w(m,k,l,n)\,,
~~~~~w(m,k,l,n).A^*=q^{\frac{(\de-1)(m-k+l-n)}s}w(m,k,l,n)\,,
$$
$$
w(m,k,l,n).B=q^{\frac{m+k-l-n+1}2-\de(n-1)}\frac{1-q^{2n}}{1-
q^2}w(m,k+1,l,n-1)+
$$
\beq{act}
q^{\frac{m+k-5l+3n+5}2-\de(k-l+n+1)}\frac{1-q^{2l}}{1-q^2}w(m+1,
k,l-1,n)\,,
\eq
$$
w(m,k,l,n).C=q^{\frac{m-3k-l-n+3}2+\de
n}\frac{1-q^{2k}}{1-q^2}w(m,k-1,l,n+1)
$$
$$
+q^{\frac{-3m-3k+3l-n+3}2+\de(k-l+n)}\frac{1-q^{2m}}{1-q^2}w(m-
1,k,l+1,n)\,.
$$
Introduce the group-like operator $M$ that acts on $\mM_{\de,q}^{1,3}$ as
\beq{3.9}
\mat{X_1}{X_2}{X_3}{X_4}.M=
\mat{q^{\frac12} X_1}{q^{\frac12} X_2}
{q^{\frac12}X_3}{q^{\frac12}X_4}\,.
\eq
It has the following properties
$$
M^*=M, ~~~\D(M)=M\otimes M, ~~~\epsilon(M)=1, ~~~S(M)=M^{-1}.
$$
Evidently, $M$ commutes with $A, B, C$ T¨ $A^*$.
Define the Hopf algebra ${\cal U}_q(\GL)$ generated by $A, B, C$ and $M$.
It is the quantum deformation of the classical algebra
$\GL$.
Let
\beq{3.11a}
\Om_{q,M}:=\frac{(q^{-\frac12}M^{-1}-q^{\frac12}M)^2}{(q^{-1}-q)^2}\,.
\eq
Consider the following Casimir element of ${\cal U}_q(\GL)$
\beq{3.12}
\Delta_q=[\Om_{q}-\Om_{q,M}]R^{-2}.
\eq
This operator is the quantum analog of the Laplace operator $\Delta$
(\ref{2.7}).

Define the partial differentiation acting on $w(m,k,l,n)$ in such a way
that
it does not break the ordering.
It means, in particular, that the differentiation of the ordered monomial
with respect, for example, $X_1$ takes the form
$$
D_{X_1}w(m,k,l,n)=\frac{1-q^{2k}}{1-q^2}w(m,k-1,l,n)\,.
$$
Let $T_Xf(X)=f(qX)$.

\begin{predl}
The action of the Laplace operator on ${\mM}^{1,3}_{\de,q}$ assumes the
form
$$
f(X_3,X_1,X_4,X_2).\Delta_q=\left\{\frac1{(q-q^{-1})^2}
[q^{-1}(T^{-1}_{X_1}T^{-1}_{X_2}T_{X_3}T^{-1}_{X_4}
-T^{-1}_{X_1}T_{X_2}T_{X_3}T^{-1}_{X_4}\right.
$$
$$
+T^{-1}_{X_1}T_{X_2}T_{X_3}^{-1}T^{-1}_{X_4}
-T^{-1}_{X_1}T^{-1}_{X_2}T^{-1}_{X_3}T^{-1}_{X_4})
$$
$$
-q(T^{-1}_{X_1}T_{X_2}T_{X_3}T_{X_4}+T_{X_1}T_{X_2}T_{X_3}T_{X_4}+
T_{X_1}T_{X_2}T_{X_3}T^{-1}_{X_4}-T^{-1}_{X_1}T_{X_2}T_{X_3}T^{-1}_{X_4})]
$$
\beq{Lo}
+q^{1+\de}T^{-k(-\de+1)}_{X_1}T^{-1}_{X_2}T^{-1}_{X_3}T^{l(1-\de)}_{X_4}
D_{X_2}D_{X_3}
\eq
$$
\left.
+q^{5-\de}T^{-k(\de+1)}_{X_1}T_{X_2}T_{X_3}T^{-l(3-\de)}_{X_4}D_{X_1}D_{X_4}
\right\}f(X_3,X_1,X_4,X_2)\,.
$$
\end{predl}
{\sl Proof}\\
One can define the action of the Casimir element on
${\cal U}^{(s)}_q(\SL)$ (\ref{3.6}) on the ordered monomials using
(\ref{act}).
Since
$$
w(m,k,l,n).M=q^{\frac{m+k+l+n}2}w(m,k,l,n)\,,
$$
we have from (\ref{3.12})
$$
w(m,k,l,n).\Delta_q=$$
$$
\frac1{(1-q^2)^2}\left[q^{m-k-l-n+1}+q^{m-k+l+n+3}+
q^{m+k-l+n+3}+q^{-m-k-l+n+1}\right.
$$
$$
-\left.q^{-m-k-l-n+1}-q^{m+k+l+n+3}-
q^{m-k-l+n+1}-q^{m-k-l+n+3}\right]w(m,k,l,n)
$$
$$
+q^{-m-k+l-n+1+\de(k-l+1)}
\frac{(1-q^{2m})(1-q^{2n})}{(1-q^2)^2}w(m-1,k+1,l+1,n-1)
$$
$$
+q^{m-k-3l+n+5-\de(k-l+1)}
\frac{(1-q^{2k})(1-q^{2l})}{(1-q^2)^2}w(m+1,k-1,l-1,n+1)\,.
$$
This relation implies (\ref{Lo}). $\Box$

\bigskip
\begin{rem}
In the classical limit $\lim_{q\to 1}\Delta_q=\Delta$ (\ref{2.7}).
\end{rem}

\subsection{The Laplace operator on $\mM_{\de.q}^{1,3}$ in terms of
the horospheric generators}

Define the ordered monomial
$$
\hat w(m,k,n)=(Z^*)^mH^kZ^n\,,
$$
and let
$$
F(Z^*,H,Z)=\sum_{m,k,n}a_{m,k,n}\hat w(m,k,n)\,.
$$
Consider the action of the operator $\Delta_q$ on the Schwartz space
(\ref{Sch}).
\begin{predl}
The action of the Casimir operator $\Delta_q$ in terms of horospheric
generators takes
the form
\beq{Loo}
F(Z^*,H,Z,R).\Delta_q=\frac1{(1-q^2)^2}\left[q^{-1}T_H-q^2+qT^{-1}_H\right]
F(Z^*,H,Z,R)
\eq
$$
+\ve
q^{1-\de}D_{Z^*}D_Z \ddag H^{-2}T^{\de-1}_HF(Z^*,H,Z,R)\ddag\,.
$$
\end{predl}
{\sl Proof}\\
We have already define the action of ${\cal U}^{(s)}_q(\SL)$ on the
horospheric
generators (\ref{3.11}). Then the action of
$A, A^*, B, C$ on the ordered monomial $\hat{w}(m,k,n)$ takes the form
$$
\hat{w}(m,k,n).A=q^{-n+\frac{k}{2}}\hat{w}(m,k,n),
~~~\hat{w}(m,k,n).(A^*)^s=q^{(1-\de)(-2m+k)}\hat{w}(m,k,n)\,,
$$
\beq{7.01}
\hat{w}(m,k,n).B=q^{-n+\frac{k+1}{2}}
\frac{1-q^{2n}}{1-q^2}\hat{w}(m,k,n-1)\,,
\eq
$$
\hat{w}(m,k,n).C=q^{n-\frac{3(k-1)}2}q^{\de(k-1)}
\frac{1-q^{2m}}{1-q^2}\hat{w}(m-1,k-2,n)-
q^{-n+\frac{k+3}{2}}
\frac{1-q^{2n-2k}}{1-q^2}\hat{w}(m,k,n+1)\,.
$$
Thus, we have
\beq{7.02}
\hat{w}(m,k,n).\Om_{q}=
q^{-k+1}\frac{(1-q^{k+1})^2}{(1-q^2)^2}\hat{w}(m,k,n)+
\eq
$$
+q^{(\de-1)(k-1)}\frac{(1-q^{2m})(1-q^{2n})}{(1-q^2)^2}
\hat{w}(m-1,k-2,n-1)\,.
$$
On the other hand
\beq{7.03}
\hat{w}(m,k,n)R^\al.M=q^{\frac{\al}{2}}\hat{w}(m,k,n)R^\al\,,
\eq
and therefore
$$
\hat{w}(m,k,n)R^\al.\Om_{q,M}=\left[\frac{\al}{2}\right]^2_{q^2}
\hat{w}(m,k,n)R^\al\,.
$$
it follows from (\ref{3.12}) that
$$
\hat{w}(m,k,n)R^\al.\Delta_q=\frac{q^{-k+1}(1-q^{k+\al+1})(1-q^{k-\al+1})}
{(1-q^2)^2}\hat w(m,k,n)R^\al+
$$
$$
q^{(\de-1)(k-1)}\frac{(1-q^{2m})(1-q^{2n})}{(1-q^2)^2}
\hat{w}(m-1,k-2,n-1)R^\al\,.
$$
In this way we come to (\ref{Loo}). $\Box$
\begin{rem}
In the classical (\ref{Loo}) takes the form of the Laplace operator
in horospheric coordinates
$\lim_{q\to 1}\Delta_q=\Delta$ (\ref{2.7}).
\end{rem}

\bigskip

Our main goal is to find the eigen-functions of $\Delta_q$
\beq{qef}
F_\nu(Z^*,H,Z,R).\Delta_q=\left[\frac\nu2\right]_{q^2}^2F_\nu(Z^*,H,Z,R)\,.
\eq

These functions are expressed through the $q$-exponents (\ref{exp}) and
the three types of $q$-cylindric functions (\ref{cyl}).
For $|q|\neq 1$ they can be defined by the expansion
\beq{cf}
\bfZ_\al^{(j)}(z)=\frac1{(1-q^2)^\al\G_{q^2}(\al+1)}
\sum_{m=0}^\infty\frac{q^{(2-\de)m(m+\al)}z^{\al+2m}}
{(q^2,q^2)_m(q^{2\al+2},q^2)_m2^{\al+2m}}\,,~~j=-\frac{3}{2}\de^2+\frac{5}{2}\de+2\,,
\eq
where $\G_{q^2}(\al+1)$ is the $q^2$-$\Gamma$-function (\ref{G}).
We assume that $\frac{|z|}{2(1-q^2)}<1$ for $\de=1$.
It can be checked that $\bfZ_\al^{(j)}$ satisfies the
difference equation (\ref{cyl}).

The non-commutative analog of the horospheric elementary harmonics
(\ref{elhar}) has the following form
\begin{predl}
The basic solutions of (\ref{qef}) are defined as
\beq{qef1}
F_\nu(Z^*,H,Z,R)=\bfe(\bar{\mu} Z^*)V_\al(H)\bfe(\mu
Z)\Xi_{\nu,\al}(R)\,,~~(\ve\neq 0)\,,
\eq
where $\mu,\al\in\mC$,
$$
V_\al(H)=H^{-1}\bfZ_\al^{(j)}(2(-\ve)^\oh|\mu|q^{-\frac\de2}H^{-1})\,,
$$
$$
\Xi_{\nu,\al}(R)=
\frac1R\bfZ_\al^{(3)}(2q^{1-\frac\nu2}\frac{1-q^\nu}{1-q^2}R)\,.
$$
\end{predl}
{\sl Proof}\\
Represent the solutions in the form
\beq{7.11}
F_\nu (Z^*,H,Z,R)=V_\al(Z^*,H,Z)\Xi_{\nu,\al}(R)\,.
\eq
Substituting it in (\ref{qef}) and using the comultiplication relations
(\ref{3.4}) we find
$$
\left(V_\al(Z^*,H,Z).\Om_{q}\right)
\left(\Xi_{\nu,\al}(R).\Om_q\right)R^{-2}-
\left(V_\al(Z^*,H,Z).\Om_{q,M}\right)
\left(\Xi_{\nu,\al}(R).\Om_{q,M}\right)R^{-2}
$$
$$
-\left[\frac\nu2\right]_{q^2}^2V_\al(Z^*,H,Z)\Xi_{\nu,\al}(R)=0\,.
$$
It follows from (\ref{7.01}) - (\ref{7.03}) that it can be rewritten as
$$
\left(V_\al(Z^*,H,Z).\Om_q\right)\Xi_{\nu,\al}(R)R^{-2}-
V_\al(Z^*,H,Z)\left(\Xi_{\nu,\al}(R).\Om_{q,M}R^{-2}\right)
$$
$$
-\left[\frac\nu2\right]_{q^2}^2V_\al(Z^*,H,Z)\Xi_{\nu,\al}(R)=0\,.
$$
In this way we come to the equations
\beq{7.12}
V_\al(Z^*,H,Z).\Om_q-
\frac{q^{-\al+2}-2q^2+q^{\al+2}}{(1-q^2)^2}V_\al(Z^*,H,Z)=0\,,
\eq
and
\beq{7.13}
\Xi_{\nu,\al}(R).\Om_{q,M}+\left(\frac{q^{-\nu+2}-2q^2+q^{\nu+2}}{(1-q^2)^2}R^2-
\frac{q^{-\al+2}-2q^2+q^{\al+2}}{(1-q^2)^2}\right)\Xi_{\nu,\al}(R)=0\,.
\eq

 From (\ref{3.9}) and (\ref{3.11a}) one rewrites the equation
(\ref{7.13}) as
$$
q\Xi_{\nu,\al}(q^{-1}R)-(q^{\al+2}+q^{-\al+2})\Xi_{\nu,\al}(R)+
q^3\Xi_{\nu,\al}(qR)=q^{3-\nu}(1-q^\nu)^2\Xi_{\nu,\al}(R)\,.
$$
Put $z=2q^{-\frac\nu2}(1-q^\nu)(1-q^2)^{-1}R$. Then we come to (\ref{cyl})
with $\de=1$ and $\bfZ_{\al}^{(3)}(z)=\Xi_{\nu,\al}(R)R$.

Consider now (\ref{7.12}) and put
\beq{7.14}
\ti{V}_\al(Z^*,H,Z)=\bfe(\bar\mu Z^*)V_\al(H)\bfe(\mu Z)\,.
\eq
Assume that $\ve=\pm 1$ and
$$
V_\al(H)=\sum_{k=0}^\infty c_k\frac{(1-q^2)^{2k-2}}
{(q^2,q^2)_k(q^{2\al+2},q^2)_k}H^{-\al-2k-1}\,.
$$
Substituting this expression and (\ref{exp}) in (\ref{7.14}) we express
$\ti{V}_\al$ in terms of monomials $\hat{w}(m,k,n)$.
Using the action of $\Om_q$ on monomials (\ref{7.02})
we obtain
$$
\bfe(\bar\mu Z^*)\sum_{k=0}^\infty c_k\frac{(1-q^2)^{2k-2}}
{(q^2,q^2)_k(q^{2\al+2},q^2)_k}q^{-\al-2k+2}(1-q^2)(1-q^{2\al+2k})
H^{-\al-2k-1}\bfe(\mu Z)
$$
$$
-\ve|\mu|^2\bfe(\bar\mu Z^*)\sum_{k=0}^\infty c_k\frac{(1-q^2)^{2k}}
{(q^2,q^2)_k(q^{2\al+2},q^2)_k}q^{(\al+2k+2)(\de-1)}
H^{-\al-2k-3}\bfe(\mu Z)=0\,.
$$
Then the coefficients $c_k$ satisfy the recurrence relation
$$
c_{k+1}=-\ve\bar\mu\mu c_kq^{2\al+4k+2-\de(\al+2k+2)}\,.
$$
Taking $c_0=1$ we find
$$
c_k=(-\ve)^k|\mu|^{2k}q^{(2-\de)k(k+\al)-\de k}\,.
$$
T'hen
$$
{\cal H}_\al(H)=q^{\frac{\de\al}2}|\mu|^{-\al}\sum_{k=0}^\infty(-\ve)^k
\frac{q^{(2-\de)k(k+\al)}(1-q^2)^{2k}}
{(q^2,q^2)_k(q^{2\al+2},q^2)_k}q^{-\frac\de2(\al+2k)}
(\bar\mu\mu)^{\frac\al2+k}H^{-\al-2k-1}\,.
$$
These series coincide with (\ref{cf}) up to a constant multiplier
after the replacement $2z=(-\ve)^\oh q^{-\frac\de2}H^{-1}$. $\Box$

\bigskip
\begin{rem}
In the classical limit we come to Proposition 3.1
$$
\lim_{q\to 1}\bfe(\bar{\mu} Z^*)V_\al(H)\bfe(\mu Z)\Xi_{\nu,\al}(R)=
\exp(i\mu z+i\bmu\bz)v_\al(h)\chi_{\nu,\al}(r)\,.
$$
\end{rem}

As in the classical situation one can restrict the operator $\Delta_q$
on the non-commutative homogeneous spaces.
\begin{cor}
The restrictions of $\Delta_q$ assume the form
$$
{\bf
L}_{q,\de}\,,~(H_3)~\,:~~\Delta_q=
\frac{1}{(1-q^2)^2}[q^3T_H-2q^2+qT_H^{-1}]+q^{1-\de}D_{Z^*}D_ZH^{-2}T_H^{\de-1}\,,
$$
$$
{\bf IL}_{q,\de}\,,~(dS_3)~\,:~~\Delta_q=
\frac{1}{(1-q^2)^2}[q^3T_H-2q^2+qT_H^{-1}]-q^{1-\de}D_{Z^*}D_ZH^{-2}T_H^{\de-1}\,.
$$
\end{cor}
Then we obtain the non-commutative deformations of the classical formulas
(\ref{ef}), (\ref{ef1}).
\begin{cor}
The basic harmonics on the non-commutative $H_3$, dS$_3$ and the light-cone
$\bfC_{q,\de}^{1,3}$ are
\beq{nef}
F_\nu(\bz,h,z)=
\bfe(\mu Z^*)H^{-1}\bfZ_\al^{(j)}(2i\ve^\oh|\mu|H^{-1})\bfe(\mu Z)\,,~~
\ve=\pm 1\,,
\eq
and
\beq{nef1}
F_\nu(\bz,h,z)=\bfe(\mu Z^*)H^{\al-1}\bfe(\mu Z)\,,~~\ve=0\,.
\eq
Here $\nu^2=\al^2-1$.
\end{cor}

\subsection{The case $\mM_q^{2,2}$}

The action of ${\cal U}_q({\rm SL}(2,\mR))$ on
$\mM_q^{2,2}$ coincides with with the action of ${\cal U}_q({\rm
SL}(2,\mC))$
on $\mM_{\de.q}^{1,3}$. The only difference is that $|q|=1$ in the
former case. Then it is straightforward to repeat the calculations of
the last
two subsections.

\begin{predl}
The Laplace operator $\Delta_q$ on $\mM_q^{2,2}$ coincides with
$\Delta_q$, defined in Proposition 9.1 with $\de=1$.
\end{predl}

Consider the horospheric description of $\mM_q^{2,2}$ from 8.3.
Let
$$
F(Z_2,H,Z_1)=\sum_{m,k,n}a_{m,k,n}\hat w(m,k,n).
$$
where $\hat w(m,k,n)=Z_2^mH^kZ_1^n$ and $F(Z_2,H,Z_1)$ belongs to the
Schwartz space.
Define the action of the operator $\Delta_q$.
\begin{predl}
The action of the Casimir operator $\Delta_q$ on $\mM_q^{2,2}$
in terms of horospheric generators takes
the form
\beq{Loo1}
F(Z_2,H,Z_1,R).\Delta_q=\frac1{(1-q^2)^2}\left[q^{3}T_H-2q^2+qT^{-1}_H\right]
F(Z_2,H,Z_1,R)
\eq
$$
+\ve D_{Z_2}D_{Z_1}\ddag H^{-2}F(Z_2,H,Z_1,R) \ddag
$$
$$
-\frac1{(1-q^2)^2}\left[q^{3}T_R-2q^2+qT^{-1}_R\right]F(Z_2,H,Z_1,R)\,.
$$
\end{predl}

\bigskip
Now we investigate the basic eigen-functions
$$
F_\nu(Z_2,H,Z_1,R).\Delta_q=\left[\frac\nu2\right]_{q^2}^2F_\nu(Z_2,H,Z_1,R)\,.
$$
Again we come to the equation (\ref{cyl}) for $q$-cylindrical
functions but with $|q|=1$. The series (\ref{cf}) are ill-defined in
this case.
There exists another approach to the theory of $q$-cylindrical
functions based on integral representations for an arbitrary $j$ and
generic
$q\in\mC$ \cite{KLS}.
As above we denote these functions
$\bfZ_\al^{(j)}(z)$.
\begin{predl}
The basic solutions on $\mM_q^{2,2}$ are
\beq{qef12}
F_\nu(Z^*,H,Z,R)=\bfe(\mu_2 Z_2)V_\al(H)\bfe(\mu_1
Z_1)\Xi_{\nu,\al}(R)\,,~~(\ve\neq 0)\,,
\eq
where $\mu_{1,2}\in\mR,\,\al\in\mC$,
$$
V_\al(H)=H^{-1}\bfZ_\al^{(j)}(2(-\ve)^\oh\sqrt{\mu_1\mu_2}q^{-\oh}H^{-1})
$$
$$
\Xi_{\nu,\al}(R)=
\frac1R\bfZ_\al^{(3)}(2q^{1-\frac\nu2}\frac{1-q^\nu}{1-q^2}R)
$$
\end{predl}

\bigskip
Restricting these solutions to $R=r=const$ we define the Klein-Gordon
operator on
non-commutative AdS$^\pm_3$
$$
\Delta_q=
\frac{1}{(1-q^2)^2}[q^3T_H-2q^2+qT_H^{-1}]+\ve D_{Z_2}D_{Z_1}H^{-2}\,.
$$
The basic functions on the homogeneous spaces have the form
$$
F_\nu(Z_2,H,Z_1)=
\bfe(\mu_2 Z_2)H^{-1}\bfZ_\al^{(j)}(2i\ve^\oh\sqrt{\mu_1\mu_2}
H^{-1})\bfe(\mu_1 Z_1)\,,~~\ve=\pm 1\,,
$$
$$
F_\nu(Z_2,H,Z_1)=\bfe(\mu_2 Z_2)H^{\al-1}\bfe(\mu_1 Z_1)\,,~~\ve=0\,,~~
(\nu^2=\al^2-1)\,.
$$

\section{Acknowledgments}
We are grateful to Peter Kulish and Sergey Kharchev for valuable
remarks. M.O. would like to thank
the Department of Mathematical Sciencies of the Aarchus University for
the hospitality
where the part of this work was done.
The work was supported by the grants NSh-1999-2003.2 of the scientific
schools, RFBR-03-02-17554 and CRDF RM1-2545.

\section{Appendix}
\setcounter{equation}{0}
\def\theequation{A.\arabic{equation}}

Here we reproduce some results from \cite{GR}.

{\sl The $q^2$-exponent} $e_{q^2}(x)$\\
is
\beq{exp}
e_{q^2}(x)=\frac1{(x,q^2)_\infty}=\sum_{n=0}^\infty\frac{x^n}{(q^2,q^2)_n}\,,
~~|x|<1\,.
\eq
it satisfies the difference equation
$$
D_xe_{q^2}(\mu x)=\mu\bfe_{q^2}(\mu x)\,,
$$
where
\beq{dif}
D_xf(x)=\frac{f(x)-f(q^2x)}{1-q^2}x^{-1}\,.
\eq

{\sl The $q^2$ Gamma-function}\\
\beq{G}
\G_{q^2}(x)=\frac{(q^2;q^2)_\infty}{(q^{2x};q^2)_\infty}(1-q^2)^{1-x}\,.
\eq

{\sl The $q$-cylindric functions} $\bfZ_\al^{(j)}$\\
are solutions of the
difference equation
\beq{cyl}
\bfZ_\al^{(j)}(q^{-1}z)-(q^{-\al}+q^\al)\bfZ_\al^{(j)}(z)+
\bfZ_\al^{(j)}(qz)=q^{-\de}(1-q^2)^2z^2\bfZ_\al^{(j)}(q^{1-\de}z)\,.
\eq
In \cite{GR} the $q$-cylindric functions are defined for $j=1,2$.
This equation is the second order difference equation for
$j=1,2,3$. There exist the analytic continuation of $\bfZ_\al^{(j)}$ on
the complex plane
$j\,(\de)$ \cite{KLS}.

{\sl The Jackson integral}\\
is the series
\beq{J}
\langle f\rangle=\int d_{q^2}uf(u)=
(1-q^2)^2\sum_{m\in\mZ}q^{2m}[f(q^{2m})+f(-q^{2m})]\,.
\eq

\small{

\end{document}
\begin{thebibliography}{60}

\bibitem{DN}
M.R.Douglas, N.A.Nekrasov,
{\em Rev. Mod. Physics}, {\bf 73} (2001), 977, [hep-th/0106048]

\bibitem{Sz}
R.J.Szabo,
{\em Phys. Rep.}, {\bf 378} (2003) 207, [hep-th/0109162]

\bibitem{JR}
A.Jevicki, S.Ramgoolam, "Noncommutative gravity from ads/cft
correspondence", {\em JHEP} {\bf 04} (1999) 032 [hep-th/9902059]

\bibitem{P}
P.Pouliot, " Finite number of states, de sitter space and quantum groups at
roots at unity", {\em Class. Quant. Grav.} {\bf 21} (2004) 145-162, [hep-th
0306261]

\bibitem{L}
A.Guijosa, D.Lowe, "A new twist on ds/cft", {\em Phys. Rev.} {\bf D69}
(2004), [hep-th/0312282];\\
D.Lowe "Deformed de Sitter/Conformal Field Theory Correspondence",
[hep-th/0407188]

\bibitem{LT}
Jerzy Lukierski, Henri Ruegg, Anatol Nowicki, Valerii
N. Tolstoi, "Q Deformation of Poincare Algebra",
{\em Phys.Lett.} {\bf B264} (1991) 331-338

\bibitem{W}
U. Carow-Watamura, M. Schlieker, M. Scholl, S. Watamura,
"Tensor Representation of the Quantum group SL-Q(2,C) and
Quantum Minkowski Space",
{\em Z.Phys.} {\bf C48}, (1990) 159-166

\bibitem{D}
V.K. Dobrev,
"New $q$ - Minkowski space-time and $q$ - Maxwell equations
hierarchy from $q$ - conformal invariance",
{\em Phys. Lett.} {\bf 341B} (1994) 133-138

\bibitem{Z}
S.Zakrzewski, "Poisson structures on the
Poincare group", [q-alg/9602001]

\bibitem{AR}
J.A. de Azcarraga, F. Rodenas, "Deformed Minkowski
Spasecs: Classification and Properties",
{\em J.Phys.} {\bf A29} (1996) 1215-1226,
[Q-ALG 9510011]

\bibitem{AKR}
J.A. de Azcarraga, P.P. Kulish, F. Rodenas, "On the physical Content of
Q-deformed Minkowski spaces",
{\em Phys.Lett.} {\bf B351}, (1995) 123-130,
[HEP-TH 9411121]

\bibitem{K}
P.P. Kulish, "Representations of q-Minkowski space algebra",
{\em Alg. and Anal.} {\bf 6} (1994), 195-205;
{\em St. Petersburg Math.J.} (1995) 6365-374,
[HEP-TH 9312139]

\bibitem{KR}
P.Kulish, N.Reshetikhin, "Quantum linear problem for the sin-Gordon
equation
and the higher representations", {\em LOMI notes}, {\bf 101} (1981) 101-110

\bibitem{Sck}
E.Sklyanin,
"About algebra, generated by quadratic relations",
{\em UMN}, {\bf 40} (1985) 214

\bibitem{KLS}
S. Kharchev, D. Lebedev, M. Semenov-Tian-Shansky,
"Unitary representations of $U_{q}(sl(2,R))$, the modular double, and the
multiparticle q-deformed Toda chains",
{\em Commun.Math.Phys.},{\bf 225} (2002) 573-609, [hep-th/0102180]


\bibitem{V}\ L.L.Vaksman and L.I.Korogodsky,
"Harmonic analysis on quantum hyperboloids",
Preprint ITPh-90-27P, Kiev (1990)


\bibitem{OR1}
M.A. Olshanetsky, V.-B.K. Rogov
"Unitary Representations of Quantum Lorentz Group and Quantum
Relativistic Toda
Chain", TMPh {\bf 130} (2002) 355-382,
[math.QA/0110182]


\bibitem{GR}
G.Gasper, M.Rahman
"Basic Hypergeometric series", Cambridge University Press (1990)

\end{thebibliography}
